\documentclass[journal]{IEEEtran}

\IEEEoverridecommandlockouts                              

\usepackage{multicol}
\usepackage{graphicx}
\usepackage{etoolbox}
\usepackage{amsmath} 
\usepackage{cite}
\usepackage[table,x11names]{xcolor}
\usepackage{soul,color}
\soulregister\cite7
\soulregister\ref7
\soulregister\pageref7
\usepackage{subcaption}
\usepackage[font=footnotesize]{caption}
\usepackage{epstopdf}
\DeclareGraphicsExtensions{.eps}

\begin{document}
\title{Refined Composite Multiscale Dispersion Entropy and its Application to Biomedical Signals}

\author{Hamed Azami$^{1,*}$,~\IEEEmembership{Student Member,~IEEE,}
	Mostafa Rostaghi$^{2}$, Daniel Ab{\'a}solo$^{3}$,~\IEEEmembership{Member,~IEEE},
	and~Javier Escudero$^{1}$,~\IEEEmembership{Member,~IEEE}
	\thanks{$^{1}$H. Azami and J. Escudero are with the Institute for Digital Communications, School of Engineering, The University of Edinburgh, Edinburgh, King's Buildings, EH9 3FB, United Kingdom (Phone: +44748 1478684). \quad \quad * Corresponding author, email: hamed.azami@ed.ac.uk.}
	\thanks{$^{2}$M. Rostaghi is with the Department of Mechanical Engineering, Shahid Rajaee Teacher Training University, Tehran, Iran.}
	\thanks{$^{3}$D. Ab{\'a}solo is with the Centre for Biomedical Engineering, Department of Mechanical Engineering Sciences, Faculty of Engineering and Physical Sciences, University of Surrey, Guildford, GU2 7XH, UK.}}

\maketitle

\begin{abstract}
\textit{Objective}: We propose a novel complexity measure to overcome the deficiencies of the widespread and powerful multiscale entropy (MSE), including, MSE values may be undefined for short signals, and MSE is slow for real-time applications.

\textit{Methods}: We introduce multiscale dispersion entropy (DisEn - MDE) as a very fast and powerful method to quantify the complexity of signals. MDE is based on our recently developed DisEn, which has a computation cost of O(\textit{N}), compared with O(\textit{N\textsuperscript{2}}) for sample entropy used in MSE. We also propose the refined composite MDE (RCMDE) to improve the stability of MDE. 

\textit{Results}: We evaluate MDE, RCMDE, and refined composite MSE (RCMSE) on synthetic signals and three biomedical datasets. The MDE, RCMDE, and RCMSE methods show similar results, although the MDE and RCMDE are faster, lead to more stable results, and discriminate different types of physiological signals better than MSE and RCMSE.

\textit{Conclusion}: For noisy short and long time series, MDE and RCMDE are noticeably more stable than MSE and RCMSE, respectively. For short signals, MDE and RCMDE, unlike MSE and RCMSE, do not lead to undefined values. The proposed MDE and RCMDE are significantly faster than MSE and RCMSE, especially for long signals, and lead to larger differences between physiological conditions known to alter the complexity of the physiological recordings.

\textit{Significance}: MDE and RCMDE are expected to be useful for the analysis of physiological signals thanks to their ability to distinguish different types of dynamics. The Matlab codes used in this paper are freely available at http://dx.doi.org/10.7488/ds/1982.
\end{abstract}

\begin{IEEEkeywords}
Complexity, multiscale dispersion entropy, non-linearity, biomedical signals, electroencephalogram, blood pressure.
\end{IEEEkeywords}

\IEEEpeerreviewmaketitle

\section{Introduction}
\IEEEPARstart{E}{ntropy} is an effective and broadly used measure of the irregularity and uncertainty of time series \cite{rostaghi2016dispersion,richman2000physiological}. Higher entropy shows higher uncertainty, while lower entropy corresponds to less irregularity or uncertainty of a signal \cite{sanei2007eeg}. When dealing with biomedical signals, two of the most common entropy estimators are sample entropy (SampEn) \cite{richman2000physiological} and permutation entropy (PerEn) \cite{bandt2002permutation}.  

SampEn denotes the negative natural logarithm of the conditional probability that a signal of length $N$, having repeated itself within a tolerance $r$ for $m$ sample points, will also repeat for $m+1$ sample points \cite{richman2000physiological}. For more information about SampEn, please refer to \cite{richman2000physiological}. In spite of its advantages over other entropy methods, SampEn has a computation cost of O($N^{2}$) \cite{jiang2011fast,wu2016refined}. PerEn, as a fast and powerful symbolization method, is based on the permutation patterns or the order relations of the amplitude values of a signal \cite{bandt2002permutation}. PerEn has been used in many signal processing applications, because it is computationally quick and has a computation cost of O(\textit{N}) \cite{wu2016refined}. Nevertheless, it has two key deficiencies: 1) when a time series is symbolized using the Bandt-Pompe algorithm, only the order of the amplitude values is considered and some information about the amplitude values is ignored, and 2) the impact of equal amplitude values in each embedding vector was not addressed in PerEn \cite{azami2016amplitude}. While modified PerEn (MPerEn) has been recently introduced \cite{bian2012modified} to address point 2) and weighted PerEn (WPerEn) \cite{fadlallah2013weighted} to address 1), none of them addresses both shortcomings.

To alleviate the deficiencies of PerEn, WPerEn, MPerEn, and SampEn, we proposed a new entropy method, named dispersion entropy (DisEn) \cite{rostaghi2016dispersion}. DisEn needs to neither sort the amplitude values of each embedding vector nor calculate every distance between any two composite delay vectors with embedding dimensions $m$ and $m+1$. This makes DisEn significantly faster than PerEn, WPerEn, MPerEn, and SampEn, and it leads to a computation cost of O(\textit{N}). DisEn overcomes the problem of equal values for embedding vectors and discarding some information with regard to the amplitudes for PerEn \cite{rostaghi2016dispersion,azami2016amplitude}. Finally, unlike PerEn, MPerEn, WPerEn, and even SampEn, DisEn is relatively insensitive to noise, because a small change in amplitude value will not vary its class label\cite{rostaghi2016dispersion}. The results demonstrated that DisEn, unlike PerEn, WPerEn, and MPerEn, is sensitive to changes in simultaneous frequency and amplitude values and bandwidth of signals \cite{rostaghi2016dispersion}. We also showed that DisEn outperformed PerEn in the discrimination of diverse biomedical datasets \cite{rostaghi2016dispersion,azami2016dispersion}.

Nevertheless, existing entropy methods (such as SampEn, PerEn, and DisEn) used to quantify the uncertainty of signals on a single scale \cite{shannon2001mathematical,rostaghi2016dispersion} by assessing repetitive patterns, return maximum values for completely random processes \cite{fogedby1992phase,costa2005multiscale,zhang1991complexity}. Signals recorded from subjects suffering from different pathologies usually show a more regular behavior and are associated with smaller entropy values in comparison with healthy ones \cite{costa2005multiscale}. In contrast, certain pathologies, such as cardiac arrhythmias, are associated with highly erratic fluctuations with statistical characteristics resembling uncorrelated noise. The entropy values of these noisy signals are higher than those of healthy individuals, even though the healthy individuals' time series show more physiologically complex adaptive behavior \cite{hayano1997spectral,costa2005multiscale}. To provide a unified framework for the evaluation of impact of diseases in physiological signals, multiscale entropy (MSE) \cite{costa2005multiscale} was proposed to quantify the complexity of signals over multiple temporal scales.

The complexity concept stands for “meaningful structural richness”, which may be in contrast with regularity measures defined from classical entropy approaches such as SampEn, PerEn, and DisEn \cite{bar1997dynamics,costa2005multiscale,fogedby1992phase}. In fact, a completely ordered signal with a small entropy value or a completely disordered signal with maximum entropy value is the least complex \cite{fogedby1992phase,costa2005multiscale,silva2015multiscale}. For example, white Gaussian noise (WGN) is more irregular than $1/f$ noise although the latter is more complex because $1/f$ noise contains long-range correlations and its $1/f$ decay produces a fractal structure in time \cite{fogedby1992phase,costa2005multiscale,silva2015multiscale}. The neural networks in the brain, with their structure intermediate between order and randomness are considered as another example of complexity in the physiological area \cite{stam2005nonlinear}.

In brief, the concept of complexity builds on three hypotheses: I) the complexity of a biological or physiological time series indicates its ability to adapt and function in ever-changing environment; II) a biological time series requires to operate across multiple temporal and spatial scales and so, its complexity is similarly multiscaled and hierarchical; and III) a wide class of disease states, in addition to aging, decrease the adaptive capacity of the individual, thus reducing the information carried by output variables. Therefore, the MSE focuses on quantifying the information expressed by the physiologic dynamics over multiple temporal scales \cite{costa2005multiscale,fogedby1992phase}.

The MSE has been widely used in different research fields, including biomedical applications \cite{humeau2015multiscale}. MSE has been successfully employed to characterize different pathological states, such as to diagnose depression using physiological signals, including heart rate, speech recordings, and electroencephalograms (EEGs) \cite{goldberger2012complexity}, to detect Parkinson's disease using EEGs\cite{chung2013multiscale}, and to characterize Alzheimer's disease (AD) \cite{escudero2015multiscale}. The assessment of entropy at multiple scale factors, proposed by Costa, \textit{et al}., has also inspired researchers to introduce other complexity metrics such as multiscale PerEn (MPE) \cite{morabito2012multivariate}. However, MPE, though fast, does not fulfil the key hypotheses of the concept of complexity \cite{wu2016refined}.

To increase the accuracy of entropy estimation and decrease the probability of facing situations where entropy is undefined, especially for short signals, refined composite MSE (RCMSE) was proposed \cite{wu2014analysis}. Although the RCMSE reduces the sensitivity of the MSE to the length of signals, the problem of undefined MSE and RCMSE values is still present, as shown later. In brief, MSE and RCMSE, though powerful, have the following shortcomings: (i) MSE and RCMSE values are undefined for short signals, (ii) MSE and RCMSE are not stable enough, especially for short signals, and (iii) the computation of MSE and RCMSE is not quick enough for some applications and their computational cost is O($N^{2}$) \cite{wu2016refined}. 
	
To alleviate these deficiencies, we introduce multiscale DisEn (MDE) and its improved version, \textit{i.e.}, refined composite MDE (RCMDE), as fast and powerful complexity estimators for real world signals. 	
	
Thus, the main contribution of this study is the proposal of MDE and RCMDE and the testing of these algorithms with both synthetic datasets and clinically relevant real-world signals: focal and non-focal EEGs \cite{andrzejak2012nonrandomness}, blood pressure recordings in Fantasia database\cite{iyengar1996age}, and resting-state EEGs activity in AD\cite{abasolo2006entropy,abasolo2008approximate}. In comparison with the existing entropy estimators, we show that: 1) MDE and RCMDE increase the reliability of the results and at the same time do not lead to undefined values for short signals, 2) their results are more stable for both short and long time series, 3) they are considerably faster, especially for long signals, and 4) they lead to larger differences between physiological conditions.

\section{Methods}
\subsection{Multiscale Dispersion Entropy (MDE)}
MDE is more than the combination of the coarse-graining \cite{costa2005multiscale} with DisEn. Instead, crucially, the mapping based on the normal cumulative distribution function (NCDF) used in the calculation of DisEn \cite{rostaghi2016dispersion} for the first temporal scale is maintained across all scales. In fact, in MDE and RCMDE, $\mu$ and $\sigma$ of NCDF are respectively set at the average and standard deviation (SD) of the original signal and they remain constant for all scale factors. This approach is similar to keeping $r$ constant fixed (usually 0.15 of the SD of the original signal) in the MSE-based algorithms \cite{costa2005multiscale}.

Assume we have a univariate signal of length $L$: $\textbf{u}=\{\textit{u}_{1}, \textit{u}_{2},..., \textit{u}_{L}\}$. In the MDE algorithm, the original signal \textbf{u} is first divided into non-overlapping segments of length  $\tau$, named scale factor. Then, the average of each segment is calculated to derive the coarse-grained signals as follows \cite{costa2005multiscale}:
\begin{equation} 
{{x}_{j}}^{(\tau )}=\frac{1}{\tau }\sum\limits_{b=(j-1)\tau +1}^{j\tau }{{{u}_{b}}}, \quad 1\le j\le \left\lfloor \frac{L}{\tau } \right\rfloor =N\
\end{equation}
Finally, the entropy value, using DisEn, is calculated for each coarse-grained signal.

The DisEn of the univariate time series of length $N$: $\textbf{x}=\{\textit{x}_{1}, \textit{x}_{2},..., \textit{x}_{N}\}$ is defined as follows:

1) First, $x_{j}(j=1,2,...,N)$ are mapped to $\textit{c}$ classes with integer indices from 1 to $\textit{c}$. To this end, the NCDF is first used to overcome the problem of assigning the majority of $x_{i}$ to only few classes in case maximum or minimum values are noticeable larger or smaller than the mean/median value of the signal. The NCDF maps $\textbf{x}$ into $\textbf{y}=\{\textit{y}_{1}, \textit{y}_{2},..., \textit{y}_{N}\}$ from 0 to 1 as follows: 
\begin{equation}
y_{j}=\frac{1}{\sigma\sqrt{2\pi}}\int\limits_{-\infty}^{x_{j}}{e}^{\frac{-(t-\mu)^2}{2\sigma^2}}\,\mathrm{d}t\end{equation}
where $\sigma$ and $\mu$ are the SD and mean of time series \textbf{x}, respectively. Then, we use a linear algorithm to assign each $y_{i}$ to an integer from 1 to $c$. To do so, for each member of the mapped signal, we use $z_{j}^{c}=\mbox{round}(c\cdot y_{j}+0.5)$, where  $z_{j}^{c}$ denotes the $j^{th}$ member of the classified time series and rounding involves either increasing or decreasing a number to the next digit \cite{rostaghi2016dispersion}. Although this part is linear, the whole mapping approach is non-linear because of the use of the NCDF. It is worth noting that other linear and non-linear mapping techniques can be used in this step.

2) Time series $\textbf{z}^{m,c}_i$ are made with embedding dimension $m$ and time delay $d$ according to $\textbf{z}^{m,c}_i=\{z^{c}_i,{z}^{c}_{i+d},...,{z}^{c}_{i+(m-1)d}\} $, $i=1,2,...,N-(m-1)d$ \cite{rostaghi2016dispersion}\cite{richman2000physiological}\cite{bandt2002permutation}. Each time series $\textbf{z}^{m,c}_i$ is mapped to a dispersion pattern ${{\pi }_{{{v}_{0}}{{v}_{1}}...{{v}_{m-1}}}}$, where $z_{i}^{c}=v_0$, $z_{i+d}^{c}=v_1$,..., $z_{i+(m-1)d}^{c}=v_{m-1}$. The number of possible dispersion patterns that can be assigned to each time series $\textbf{z}^{m,c}_i$ is equal to $c^{m}$, since the signal has $m$ members and each member can be one of the integers from 1 to $c$ \cite{rostaghi2016dispersion}. 

3) For each $c^{m}$ potential dispersion patterns ${{\pi }_{{{v}_{0}}...{{v}_{m-1}}}}$, relative frequency is obtained as follows:
\begin{equation}
\begin{split}
p({{\pi }_{{{v}_{0}}...{{v}_{m-1}}}})=\\
\frac{\#\{i\left| i\le N-(m-1)d,\mathbf{z}_{i}^{m,c}\text{ has type }{{\pi }_{{{v}_{0}}...{{v}_{m-1}}}} \right.\}}{N-(m-1)d}
\end{split}
\end{equation}

\setlength{\parindent}{0pt}%
where $\#$ means cardinality. In fact, $p({{\pi }_{{{v}_{0}}...{{v}_{m-1}}}})$ shows the number of dispersion patterns of ${{\pi }_{{{v}_{0}}...{{v}_{m-1}}}}$ that is assigned to $\textbf{z}^{m,c}_i$, divided by the total number of embedded signals with embedding dimension $m$.
\setlength{\parindent}{9pt}

4) Finally, based on the Shannon's definition of entropy, the DisEn value is calculated as follows:
\begin{equation}
\begin{split}
DisEn(\mathbf{x},m,c,d)=-\sum_{\pi=1}^{c^m} {p({{\pi}_{{{v}_{0}}...{{v}_{m-1}}}})\cdot\ln }\left(p({{\pi}_{{{v}_{0}}...{{v}_{m-1}}}}) \right)      
\end{split}
\end{equation}

When all possible dispersion patterns have equal probability value, the highest value of DisEn is obtained, which has a value of $ln ({{c}^{m}})$. In contrast, if there is only one $ p({{\pi }_{{{v}_{0}}...{{v}_{m-1}}}}) $ different from zero, which demonstrates a completely regular/predictable time series, the smallest value of DisEn is obtained \cite{rostaghi2016dispersion}. Note that we use the normalized DisEn as $\frac{DisEn}{\ln ({{c}^{m}}) }$ in this study \cite{rostaghi2016dispersion}. 

\subsection {Refined Composite Dispersion Entropy (RCMDE)}
In RCMDE, for scale factor $\tau$, $\tau$ different time series, corresponding to different starting points of the coarse graining process are created and the RCMDE value is defined as the Shannon entropy value of the averages of the dispersion patterns of those shifted sequences. The \textit{k}\textsuperscript{th} coarse-grained time series $\textbf{x}_{k}^{(\tau )}=\{{{x}_{k,1}}^{(\tau )},{{x}_{k,2}}^{(\tau )},...\}$ of \textbf{u} is as follows:
\begin{equation}
{{x}_{k,j}}^{(\tau )}=\frac{1}{\tau }\sum\limits_{b=k+\tau (j-1)}^{k+\tau j-1}{{{u}_{_{b}}}},\quad 1\le j\le N,\,\,\,  1\le k\le \tau 
\end{equation}
Then, for each scale factor, RCMDE is defined as follows:
\begin{equation}
\begin{split}
RCMDE(\mathbf{x},m,c,d,\tau)=\\
-\sum_{\pi=1}^{c^m} {\bar{p}({{\pi}_{{{v}_{0}}...{{v}_{m-1}}}})\cdot\ln }\left(\bar{p}({{\pi}_{{{v}_{0}}...{{v}_{m-1}}}}) \right)      
\end{split}
\end{equation}       
where $\bar{p}({{\pi}_{{{v}_{0}}...{{v}_{m-1}}}})=\frac{1}{\tau}\sum_{1}^{\tau}{p}_{k}^{(\tau )}$ with the relative frequency of the dispersion pattern $\pi$ in the series $\textbf{x}_{k}^{(\tau)} (1 \leq k\leq \tau)$.

\subsection{Parameters of MDE and RCMDE}
There are four parameters for MDE, including the embedding dimension \textit{m}, the number of classes \textit{c}, the time delay \textit{d}, and the maximum scale factor $\tau_{max}$.
In practice, it is recommended $ d=1 $, because aliasing may occur for $ d>1 $ \cite{rostaghi2016dispersion}. Clearly, we need $c>1$ in order to avoid the trivial case of having only one dispersion pattern. For MDE and RCMDE, here, we use $c=6$ for all signals according to \cite{rostaghi2016dispersion}, although the range $2<c<9$ leads to similar results. For more information about \textit{c}, \textit{m}, and \textit{d}, please refer to \cite{rostaghi2016dispersion}.

To work with reliable statistics to calculate DisEn, it was recommended that the number of potential dispersion patterns is smaller than the length of the signal ($c^{m}<L$) \cite{rostaghi2016dispersion}. For MDE, since the coarse-graining process causes the length of a signal decreases to $\left\lfloor \frac{L}{\tau_{max} } \right\rfloor$, it is recommended $c^{m}<\left\lfloor \frac{L}{\tau_{max} } \right\rfloor$. In RCDME, we consider $\tau$ coarse-grained time series with length  $\left\lfloor \frac{L}{\tau_{max} } \right\rfloor$. Therefore, the total sample points calculated in RCMDE is $\tau\times\left\lfloor \frac{L}{\tau_{max}} \right\rfloor\approx L$. Thus, the RCMDE follows $c^{m}<L$, leading to more reliable results, especially for short signals.

\section{Evaluation Signals}
In this section, we briefly explain the synthetic and real signals used in this study to evaluate the behaviour of MDE and RCMDE.
\subsection{Synthetic Signals}
1) The complexity of $1/f$ noise is higher than WGN, while the irregularity of the former method is lower than the latter one \cite{costa2005multiscale,fogedby1992phase}. Accordingly, WGN and $1/f$ noise are two important signals to evaluate the multiscale entropy approaches \cite{costa2005multiscale,fogedby1992phase,silva2015multiscale,wu2016refined,humeau2016refined}. For more information about WGN and $1/f$ noise, please refer to \cite{costa2005multiscale,azami2016improved}.

2) To understand the relationship between MDE, RCMDE, and RCMSE, and the level of noise affecting quasi-periodic time series, we generated an amplitude-modulated quasi-periodic signal with additive WGN of diverse power. First, we created signal as an amplitude-modulated sum of two cosine waves with frequencies at 0.5 Hz and 1 Hz. The length and the sampling frequency of the signal are 100 s and 150 Hz, respectively. The first 20 s of this series (100 s) does not have any noise. Then, WGN was added to the time series \cite{azami2016improved}.

3) To find the dependence of MDE, RCMDE, and RCMSE with changes from periodicity to non-periodic non-linearity, a logistic map is used. This analysis is relevant to the model parameter $\alpha$ as: $u_{k}=\alpha u_{k-1}(1-u_{k-1})$, where the signal $x$ was generated varying the parameter $\alpha$ from 3.5 to 3.99. When $\alpha$ is equal to 3.5, the signal oscillated among four values. For $3.5<\alpha <3.57$, the time series is periodic and the number of values doubles progressively. For $\alpha$ between 3.57 and 3.99, the time series is chaotic, although it has windows of periodic behavior (e.g., $\alpha\approx3.8$) \cite{baker1996chaotic,ferrario2006comparison,escudero2009interpretation}. Note that the signal has a length of 100 s with a sampling frequency of 150 Hz.

\subsection{Real Biomedical Datasets}
EEG and blood pressure recordings are affordable, non-invasive, and widely-used to detect different physiological states \cite{sanei2007eeg,iyengar1996age}. Using these signals, we employ RCMDE and MDE to discriminate focal signals from non-focal ones, elderly from young subjects, and AD patients from controls, as three broadly-used applications in complexity-based methods.

\textit{1) Dataset of Focal and Non-focal Brain Activity}: The ability of RCMDE and MDE to discriminate focal signals from non-focal ones is evaluated by the use of a publicly-available EEG dataset \cite{andrzejak2012nonrandomness}. The dataset includes 5 patients and, for each patient, there are 750 focal and 750 non-focal time series. The length of each signal was 20 s with sampling frequency of 512 Hz (10240 samples). For more information, please, refer to \cite{andrzejak2012nonrandomness}. Before computing the entropies, all time series were digitally band-pass filtered between 0.5 and 150 Hz using a fourth-order Butterworth filter. Backward and forward filtering was employed to minimize phase distortions \cite{andrzejak2012nonrandomness}. Finally, the time series were digitally filtered using a Hamming window FIR band-pass filter of order 200 and cut-off frequencies 0.5 Hz and 40 Hz, a band typically used in the analysis of brain activity.

\textit{2) Fantasia Dataset of Blood Pressure Recordings}: To further evaluate MDE and RCMDE, we use uncalibrated continuous non-invasive blood pressure recordings of the Fantasia database \cite{iyengar1996age}. This included 10 young (21-34 years old) and 10 old (68-85 years old) rigorously-screened healthy individuals. Each group had 5 women and 5 men. All 20 individuals remained in an inactive state in sinus rhythm when watching the movie Fantasia to help maintain wakefulness. All the recordings were digitized at 250 Hz (1,000,000 samples) \cite{iyengar1996age}. For more information, please see \cite{iyengar1996age}.

\textit{3) Surface EEG Dataset of Brain Activity in AD}: The dataset includes 11 AD patients (5 men; 6 women; age: 72.5 $\pm$ 8.3 years, all data given as mean $\pm$ SD; mini mental state examination (MMSE): 13.3 $\pm$ 5.6) and 11 age-matched control subjects (7 men; 4 women; age: 72.8 $\pm$ 6.1 years; MMSE: 30 $\pm$ 0) \cite{abasolo2006entropy}\cite{abasolo2008approximate}. The EEG signals were recorded using the international 10-20 system, in an eyes closed and resting state with a sampling frequency of 256 Hz from the Alzheimer's Patients’ Relatives Association of Valladolid (AFAVA), Spain. Informed consent was obtained for all 22 subjects and the local ethics committee approved the study. Before band-pass filtering with cut-off frequencies 0.5 and 40 Hz and a Hamming window with order 200, the signals were visually examined by an expert physician to select 5 s epochs (1280 samples) with minimal artifacts for analysis. More details can be found in \cite{abasolo2006entropy,abasolo2008approximate}.

\section{Results and Discussion}
\subsection{Synthetic Signals}
Fig. 1(a), 1(b), and 1(c) respectively show the results obtained for MDE, RCMDE, and RCMSE using 40 different WGN and $1/f$ noise signals with the length of 20,000 samples. All the results are consistent with the fact that $1/f$ noise has more complex structure than WGN, and WGN is more irregular than $1/f$ noise \cite{costa2005multiscale,silva2015multiscale,fogedby1992phase}. At short scale factors, the entropy values of WGN signals are higher than those of $1/f$ noise. However, at higher scale factors, the entropy value for the coarse-grained $1/f$ noise signal stays almost constant, while for the coarse-grained WGN signal monotonically decreases. For WGN, when the length of the signal, obtained by the coarse-graining process, decreases (i.e., the scale factor increases), the mean value of each segment converges to a constant value and the SD becomes smaller. Therefore, no new structures are revealed on higher scales. This demonstrates WGN time series contain information only in small time scales \cite{costa2005multiscale}\cite{silva2015multiscale}. For all MSE-based methods, we set $d=1$, $m=2$, and $ r=0.15$ of the SD of the original signal \cite{richman2000physiological}. Here, for WGN and $1/f$ noise, $\tau_{max}$ and $m$ respectively were 20 and 2 for MDE and RCMDE, according to Section II.C.

\begin{figure*}
	\centering
	\begin{multicols}{3}
		\includegraphics[width=6.2cm,height=3.5cm]{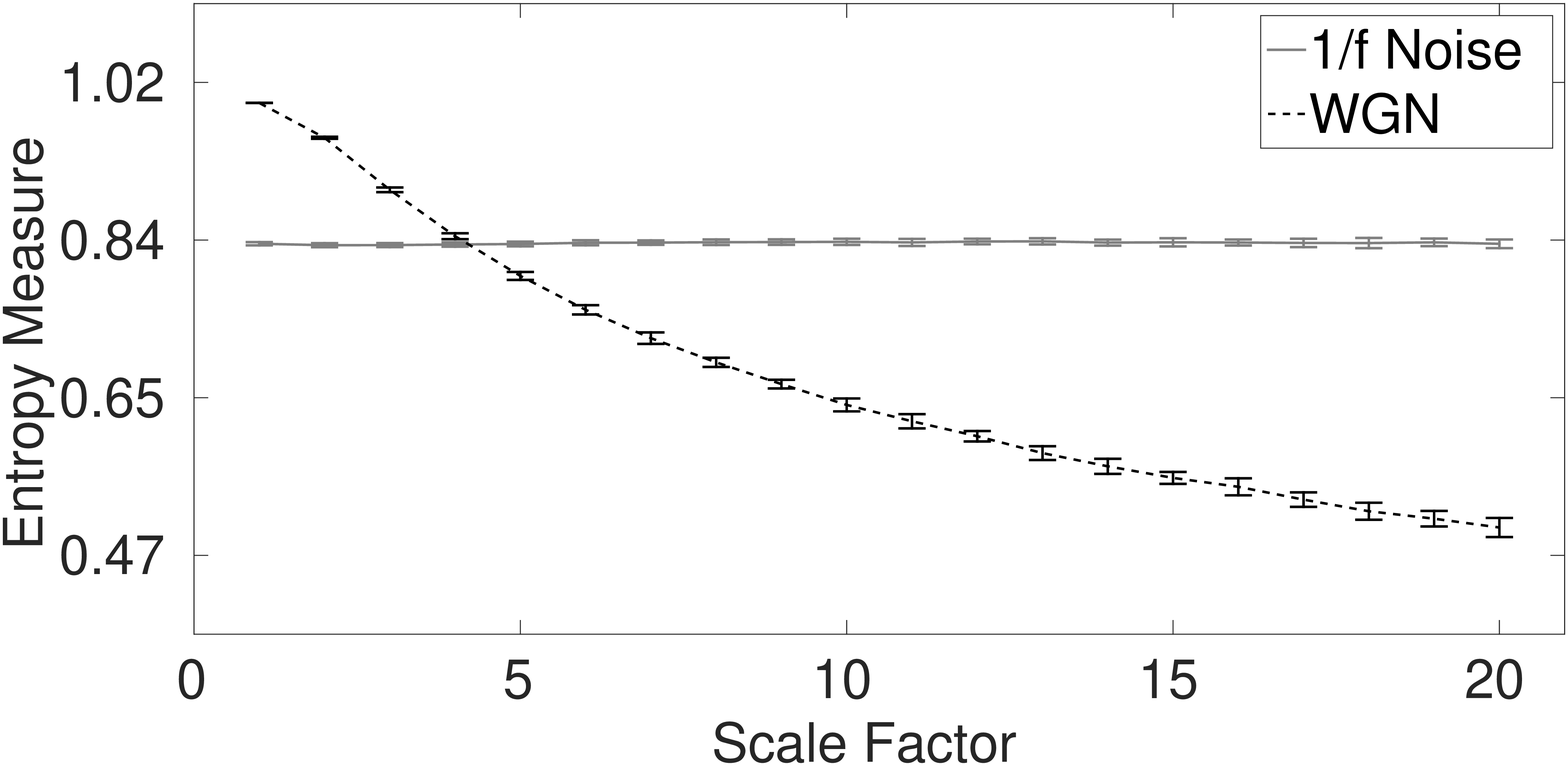}
		\footnotesize{(a) MDE}
		\includegraphics[width=6.2cm,height=3.5cm]{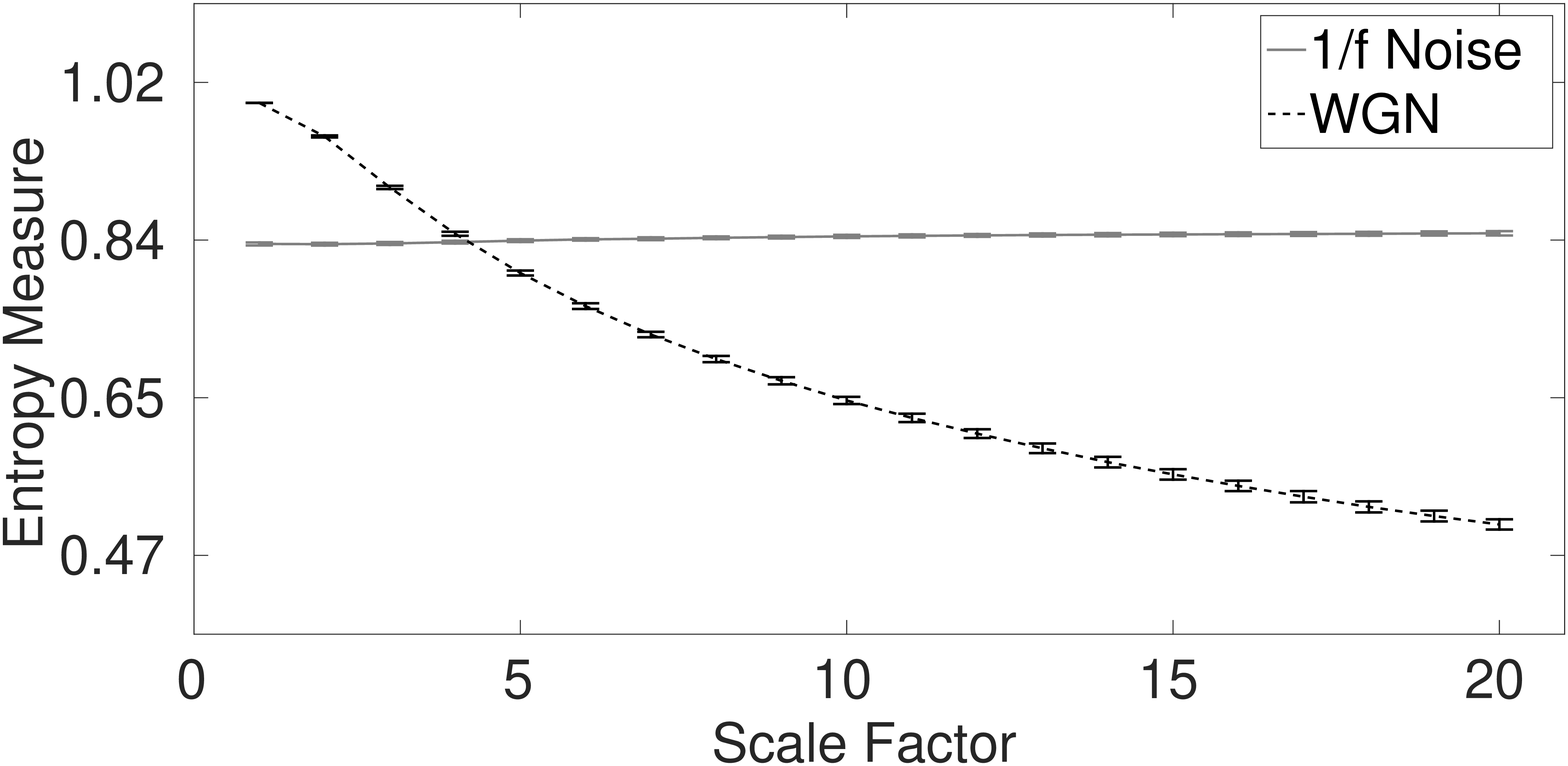}
		\footnotesize{(b) RCMDE}
		\includegraphics[width=6.2cm,height=3.5cm]{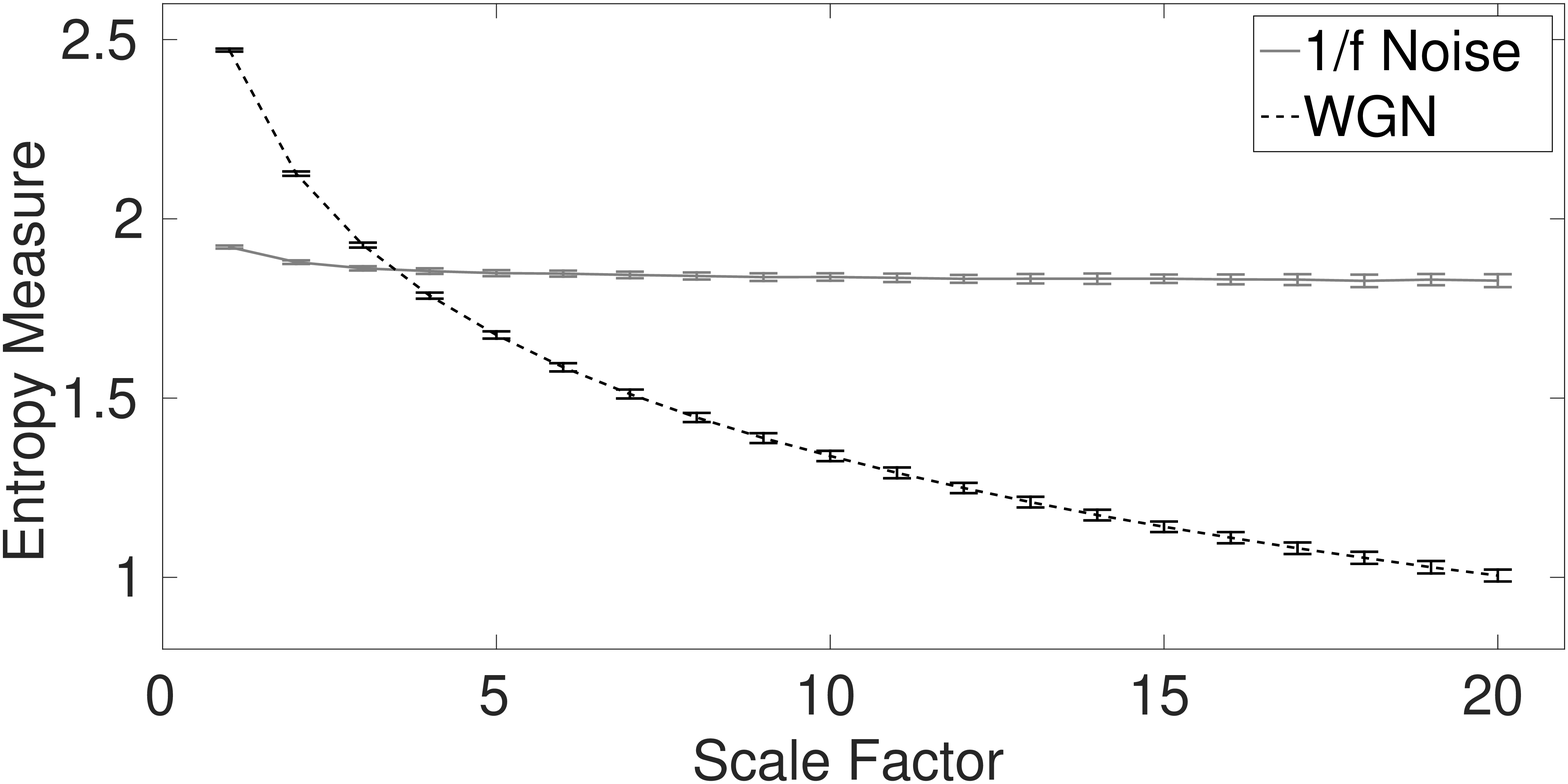}
		\footnotesize{(c) RCMSE}
	\end{multicols}
	\caption{Mean value and SD of results of the (a) MDE, (b) RCMDE, and (c) RCMSE computed from 40 different $1/f$ noise and WGN test signals.}
	\label{figurelabel}
\end{figure*}

To compare the results obtained using the MSE, RCMSE, MDE, and RCMDE, we used the coefficient of variation (CV) defined as the SD divided by the mean. We use such a measure because the SDs of signals may increase or decrease proportionally to the mean. Accordingly, the CV, as a standardization of the SD, allows comparison of variability estimates irrespective of the magnitude values. We investigate WGN and $ 1/f $ noise results at scale factor 10 as a trade-off between short and long scales. As can be seen in Table I, the RCMDE and RCMSE are more stable than MDE and RCMSE, respectively, showing the importance of the refined composite technique to improve the stability of results for noisy signals. Moreover, the CVs for MDE and RCMDE are noticeably smaller than those for MSE and RCMSE, respectively. Overall, the smallest CV values for $1/f$ noise and WGN are reached by RCMDE.

\begin{table}
	\centering
	\label{tab:table4}
	\centering     
	\caption{\footnotesize CV values of the proposed and classical multiscale entropy-based analyses at scale factor 10 for 1/$ f $ noise and WGN.}
	\centering     
	\begin{tabular}{c*{5}{c}}
		
		Time series &MSE& RCMSE& MDE& RCMDE \\
		\hline
		$ 1/f$ noise     & 0.0101& 0.0056& 0.0044& 0.0022\\
		WGN  &0.0152& 0.0087&0.0119&0.0066
	\end{tabular}
\end{table}

To evaluate the computation time of MSE (with \textit{m}=1 and 2 for completeness), MDE (\textit{m}=2 and 3, likewise), RCMSE (\textit{m}=2), and RCMDE (\textit{m}=3), we use WGN signals with different lengths, changing from 100 to 100,000 sample points. The results are shown in Table II. The simulations have been carried out using a PC with Intel (R) Xeon (R) CPU, E5420, 2.5 GHz and 8-GB RAM by MATLAB R2015a. For 100 and 300 sample points, MSE ($ m=1$ and 2) and RCMSE ($m=1$) lead to undefined values at least at several scale factors. This does not happen for MDE and RCMDE. This fact proves the superiority of MDE-based methods over MSE-based ones for short signals. There are no big differences between the computation time for the MSE with \textit{m}=1 and 2 or for the MDE with \textit{m}=2 and 3. The results show that for the different number of sample points, MDE and RCMDE are noticeably faster than MSE and RCMSE, respectively. This computational advantage of MDE and RCMDE increases notably with the signal length. It is in agreement with the fact that the computational cost of SampEn and DisEn are O($N^{2}$) \cite{wu2016refined} and O($N$), respectively \cite{rostaghi2016dispersion}. Note that the MSE and RCMSE codes used in this paper are publicly-available at http://dx.doi.org/10.7488/ds/1477.

\begin{table*}
	\label{tab:table4}\caption{The computational time of MSE, MDE, RCMSE, and RCMDE.} 
	\begin{tabular}{c*{8}{c}}
		Number of samples $\rightarrow$   &100& 300& 1,000& 3,000& 10,000& 30,000& 100,000 \\

		\hline
		MSE ($m=1$)          &undefined at all scales&undefined at several scales& 0.16 s&0.65 s& 4.08 s&25.87 s& 202.43 s \\
		MSE ($m=2$)     &undefined at all scales&undefined at all scales&undefined at several scales& 0.72 s& 4.59 s& 27.50 s&210.18 s\\
	RCMSE ($m=2$)     &undefined at all scales&undefined at several scales&0.94 s&3.33 s&16.08 s& 84.75 s&624.41 s\\
		MDE ($m=2$)          &0.02 s&0.02 s&0.05 s&0.15 s&0.45 s&1.42 s&4.52 s\\
		MDE ($m=3$)     &0.03 s&0.03 s&0.06 s&0.16 s&0.48 s&1.45 s&4.67 s\\
		RCMDE ($m=3$)     &0.09 s&0.15 s&0.40 s&1.10 s&3.49  s&10.52 s&34.15 s\\
		
	\end{tabular}
\end{table*}

The multiscale methods are applied to the logistic map and quasi-periodic signals with additive noise using a moving window of 1500 samples (10 s) with 90\% overlap. Here, for MDE and RCMDE, $\tau_{max}$ and $m$ respectively were 15 and 2, according to Section II.C. Fig. 2(a), (b), and (c) respectively show the MDE-, RCMDE- and RCMSE-based profiles using the quasi-periodic signal with increasing additive noise power. As expected, the entropy values for all three methods increase along the signal. At high scale factors, the entropy values decrease because of the filtering nature of the coarse-graining process. To sum up, the results show all the methods lead to the similar results, although the RCMDE results are slightly more stable than MDE ones, as evidenced by the smoother nature of variations in Fig. 2(b) in comparison with Fig. 2(a). Hence, when a high level of noise is present, RCMDE leads to more stable results than MDE.

The results with MDE, RCMDE, and RCMSE using a window of 10 s (1500 samples) with 90\% overlap moving along the logistic map with the parameter $\alpha$ changing linearly from 3.5 to 3.99 are shown in Fig 2(d), 2(e), and 2(f), respectively. As expected, the entropy values, obtained by the MDE, RCMDE, and RCMSE, generally increase along the signal, except for the downward spikes in the windows of periodic behavior (e.g., for $\alpha=3.8$ - shown by arrows in Fig. 2(d), 2(e), and 2(f)). This fact is in agreement with Fig. 4.10 (page 87 in \cite{baker1996chaotic}). For increasing scale factors, MDE, RCMDE, and RCMSE lead to an increase until $\tau=2$ and $\tau=3$, respectively, then a decrease. The results show that all the methods lead to similar results. In particular, there is little difference between MDE and RCMDE. Hence, when signals do not have noticeable noise, MDE and RCMDE have quite similar performance, although MDE is significantly faster due to avoiding having to repeat the coarse-graining process within each temporal scale.

\begin{figure*}
	\centering
	\begin{multicols}{3}
		\includegraphics[width=6.2cm,height=3.5cm]{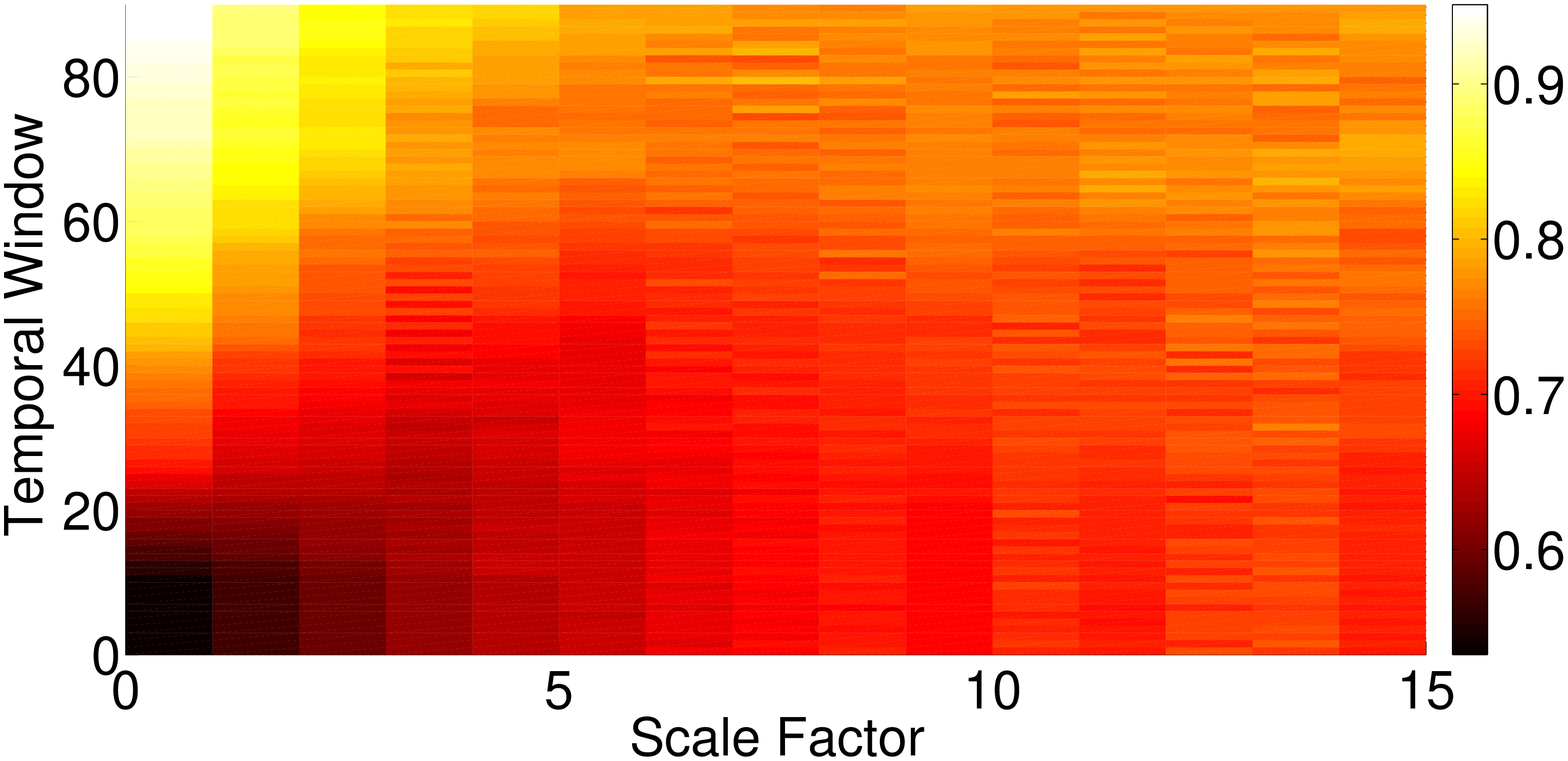}
		\footnotesize{(a) MDE results for quasi-periodic time series with increasing additive noise power}
		\includegraphics[width=6.2cm,height=3.5cm]{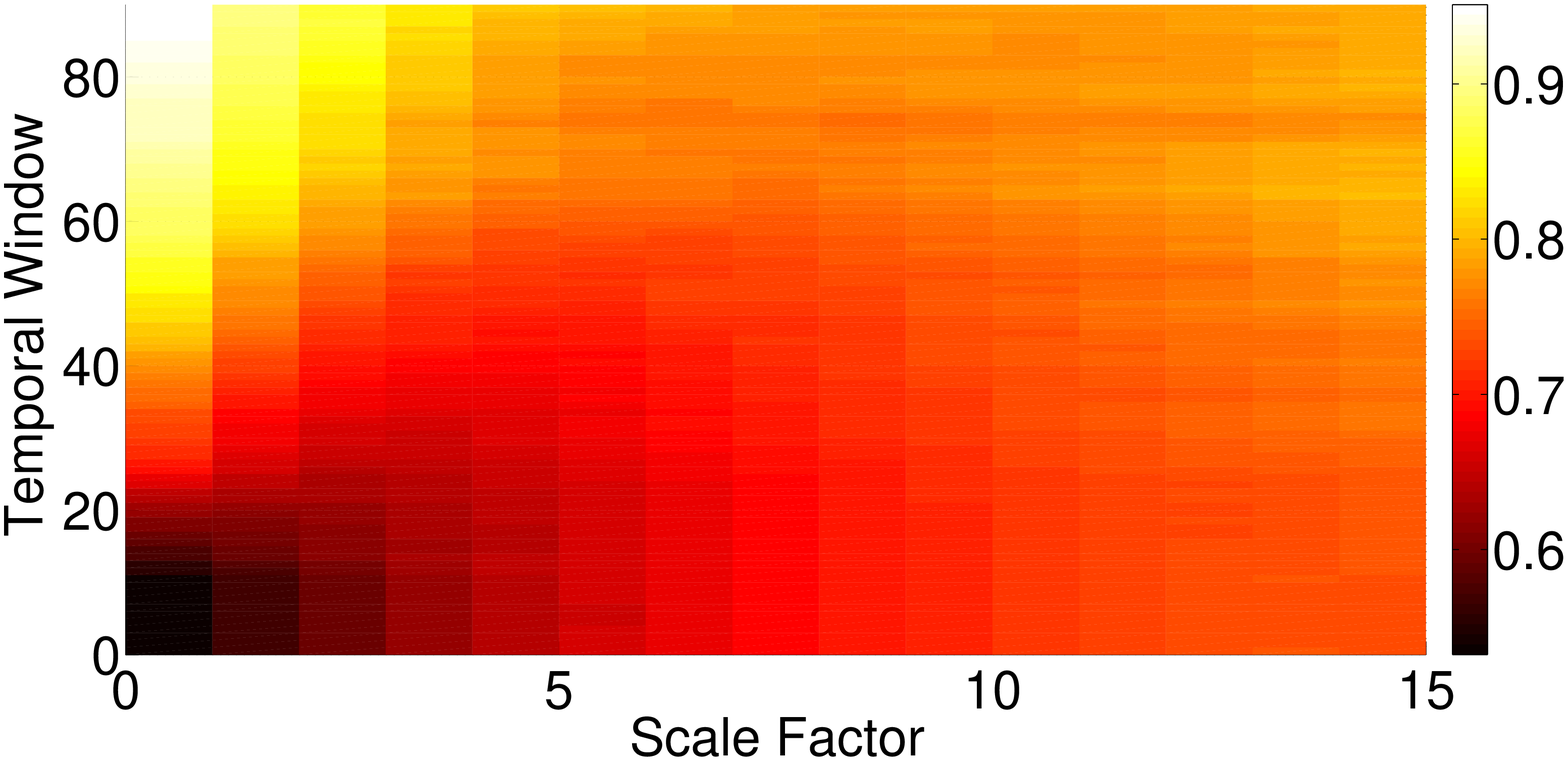}
		\footnotesize{(b) RCMDE results for quasi-periodic time series with increasing additive noise power}
		\includegraphics[width=6.2cm,height=3.5cm]{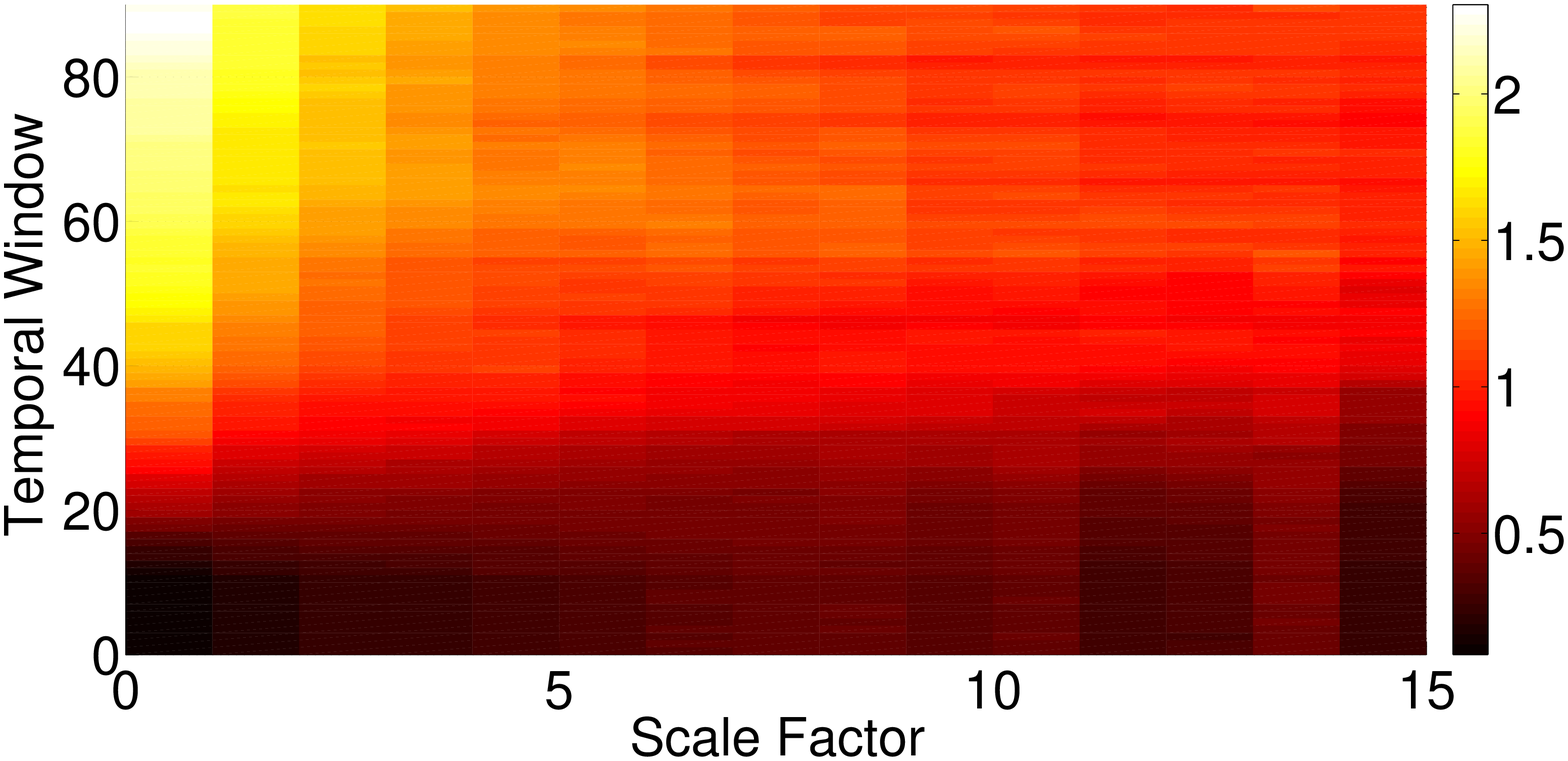}
		\footnotesize{(c) RCMSE results for quasi-periodic time series with increasing additive noise power}
	\end{multicols}
	\begin{multicols}{3}
		\centering
		\includegraphics[width=6.2cm,height=3.5cm]{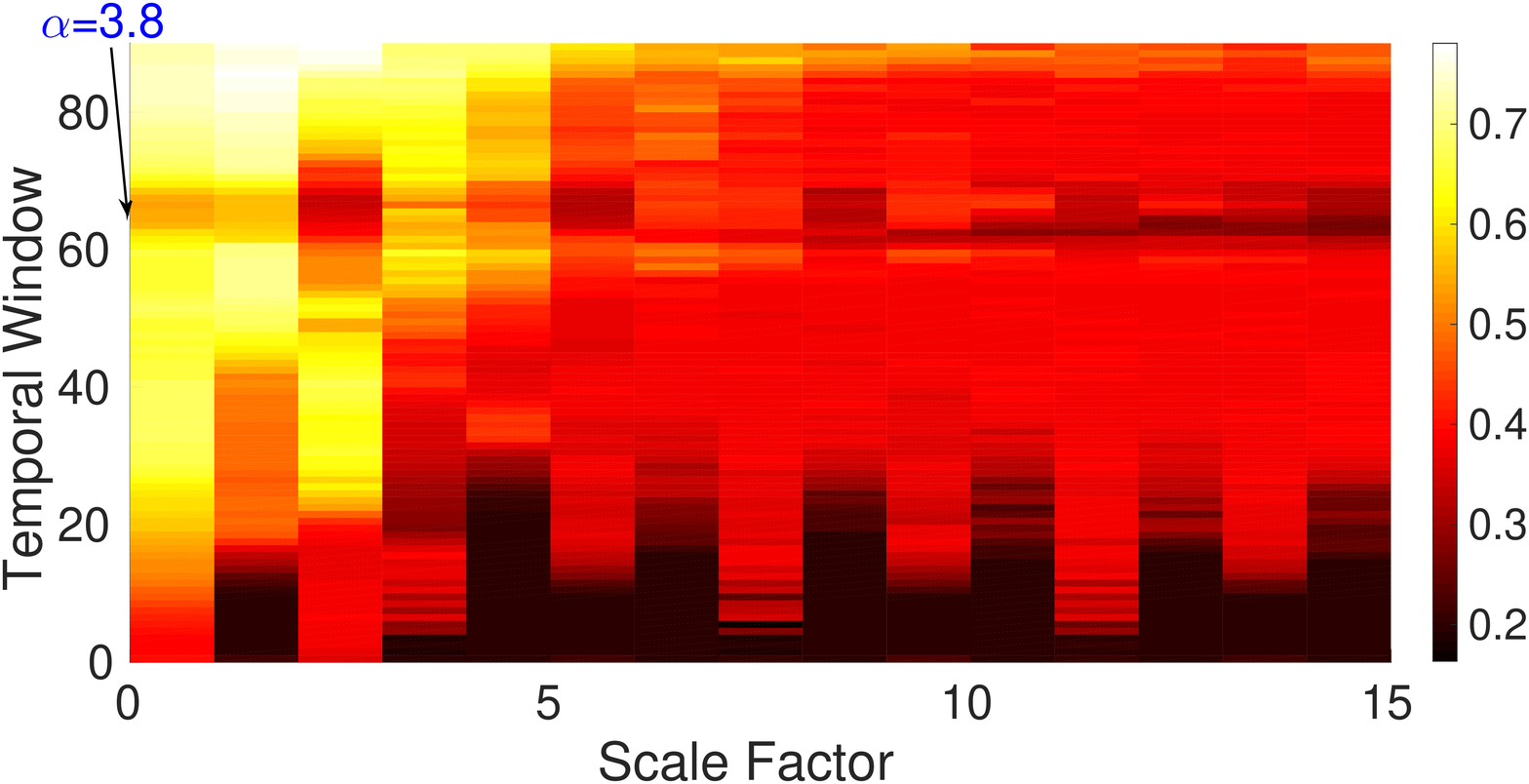}
		\footnotesize{(d) MDE results for logistic map}
		\includegraphics[width=6.2cm,height=3.5cm]{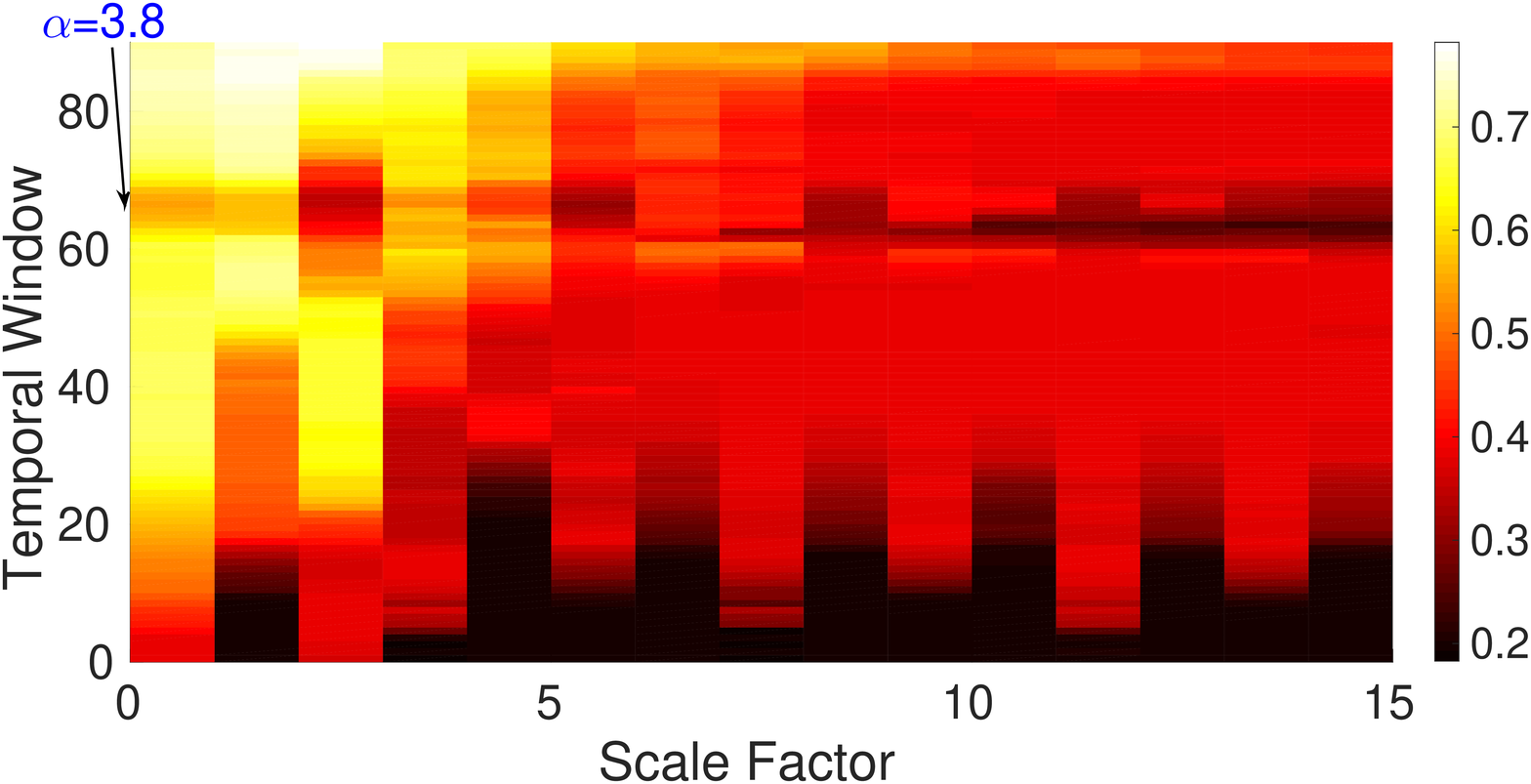}
		\footnotesize{(e) RCMDE results for logistic map}
		\includegraphics[width=6.2cm,height=3.5cm]{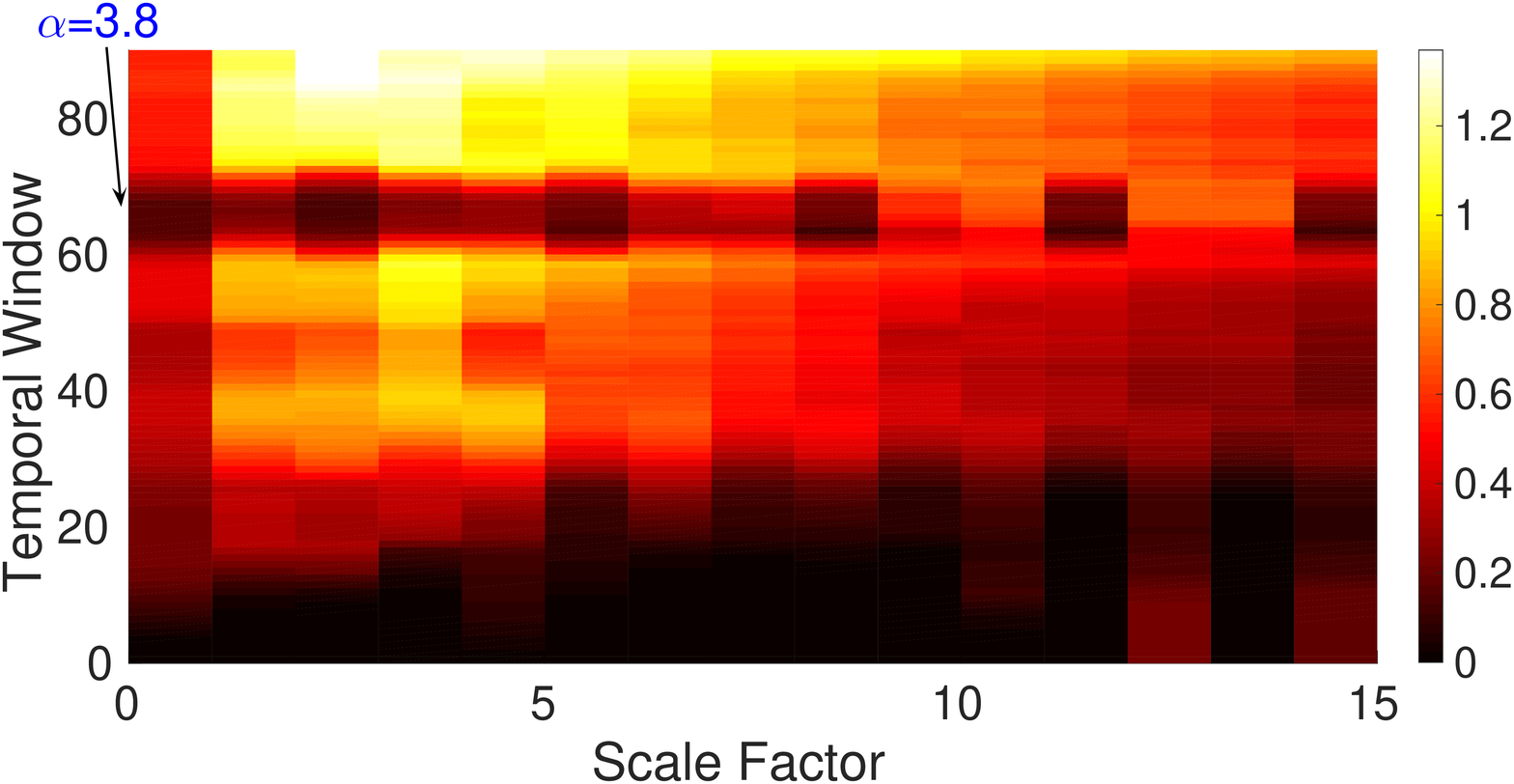}
		\footnotesize{(f) RCMSE results for logistic map}
		
	\end{multicols}
	
	\caption{Results of the tests performed to gain better understanding of (a) MDE and (b) RCMDE in comparison with (c) RCMSE using a window moving along the quasi-periodic time series with increasing additive noise power, showing an increase in entropy values along the signal (temporal window); and (d) MDE, (e) RCMDE, and (f) RCMSE using a window moving along the logistic map with varying parameter $\alpha$ from 3.5 to 3.99 showing an increase in entropy values along the signal, except for the downward spikes in the windows of periodic behavior (e.g., for $\alpha=3.8$ - shown by the arrows)}
	\label{figurelabel}

\end{figure*}

\subsection{Real Biomedical Datasets}
In the physiological complexity literature, it is hypothesized that healthy conditions correspond to more complex states due to their ability to adapt to adverse conditions, exhibiting long range correlations, and rich variability at multiple scales, while aged and diseased individuals present complexity loss. That is, they lose the capability to adapt to such adverse conditions \cite{costa2005multiscale}. Accordingly, we employ the MDE and RCMDE to characterize different kinds of biomedical signals to detect different pathological states.

\textit{1) Dataset of Focal and Non-focal Brain Activity}: For the focal and non-focal EEG dataset, the results obtained by MDE, RCMDE, and RCMSE, respectively, depicted in Fig. 3(a), (b), and (c), show that non-focal signals are more complex than focal ones. This fact is in agreement with previous studies \cite{andrzejak2012nonrandomness,sharma2015application}. The results show that the MDE, RCMDE, and RCMSE lead to similar results, albeit the MDE-based methods are significantly faster than MSE-based ones. Note that, for MDE and RCMDE, $\tau_{max}$ and $m$ respectively were 30 and 3.

\begin{figure*}
\centering
\begin{multicols}{3}
\includegraphics[width=6.2cm,height=3.3cm]{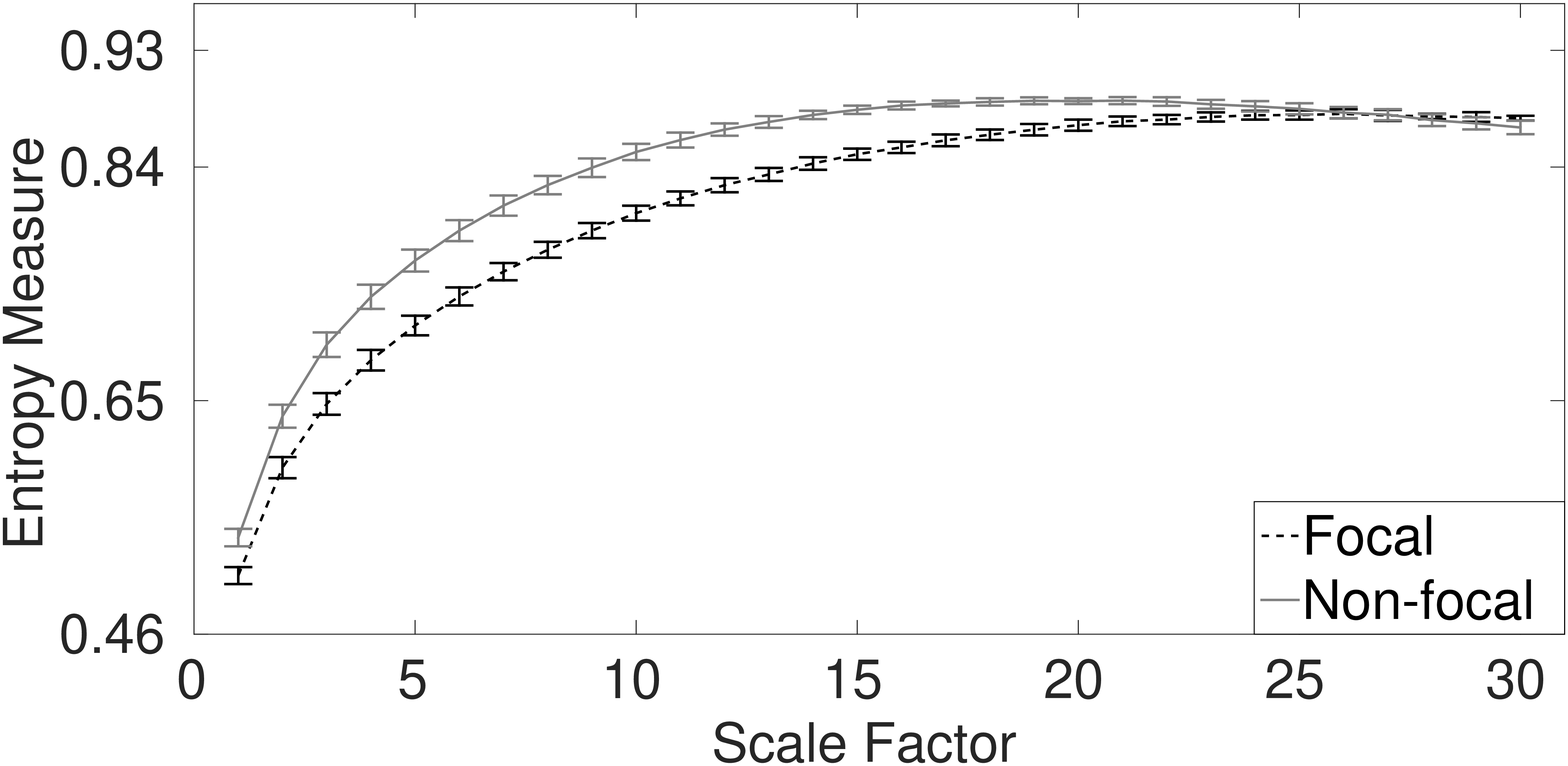}
\footnotesize{(a) MDE for focal and non-focal EEGs}
\includegraphics[width=6.2cm,height=3.3cm]{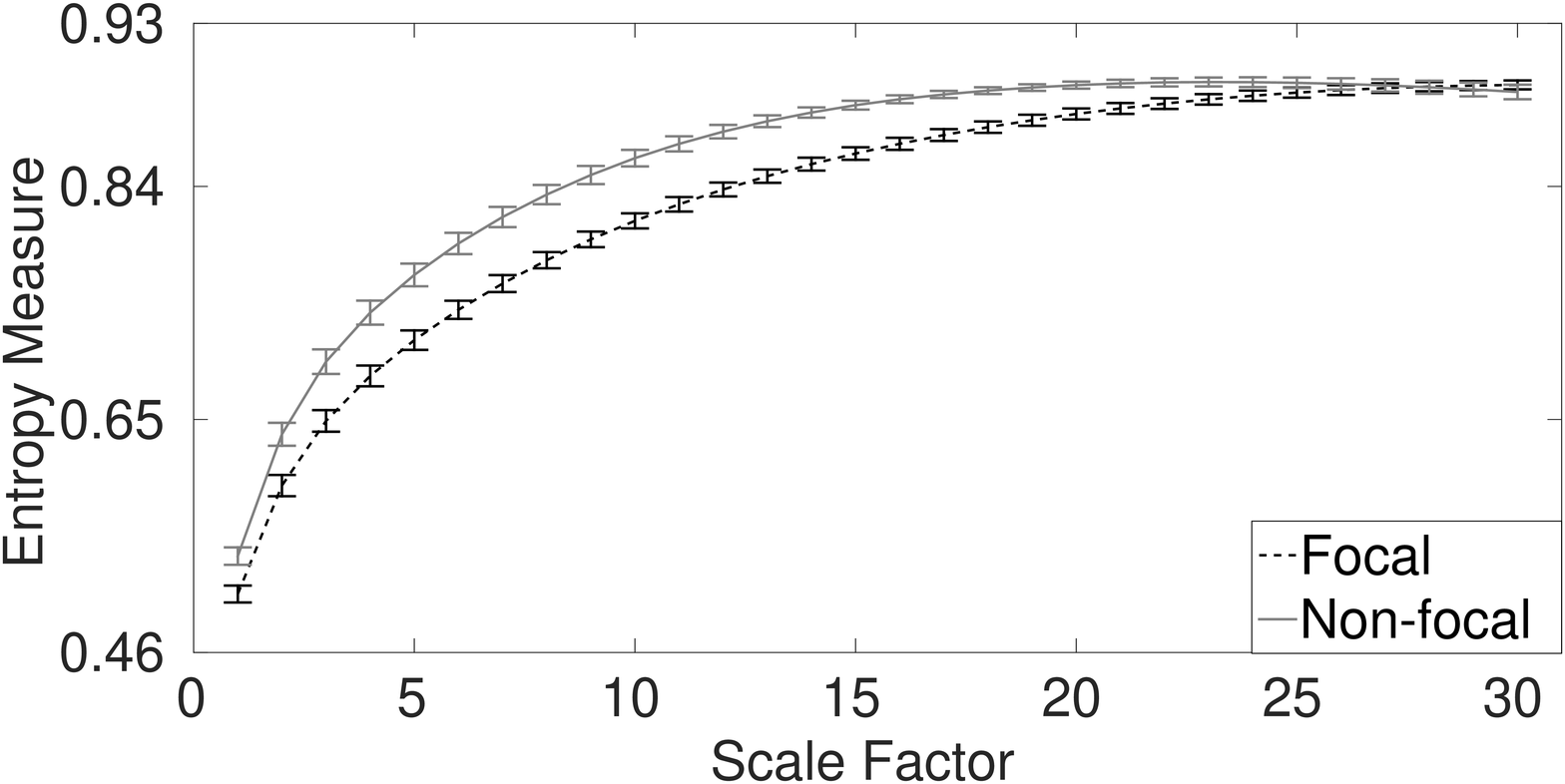}
\footnotesize{(b) RCMDE for focal and non-focal EEGs}
\includegraphics[width=6.2cm,height=3.3cm]{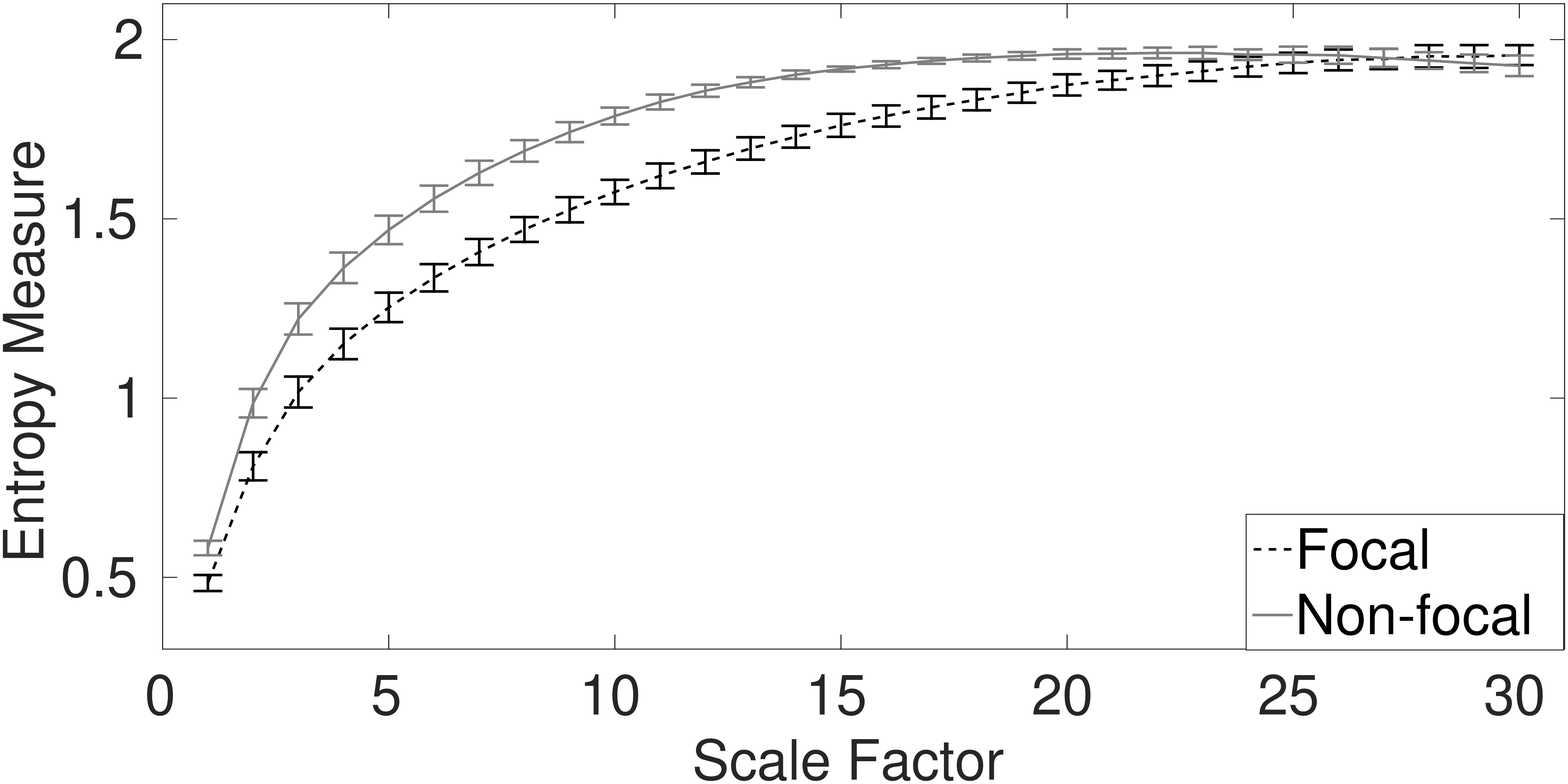}
\footnotesize{(c) RCMSE for focal and non-focal EEGs}
\end{multicols}
\begin{multicols}{3}
\centering
\includegraphics[width=6.2cm,height=3.3cm]{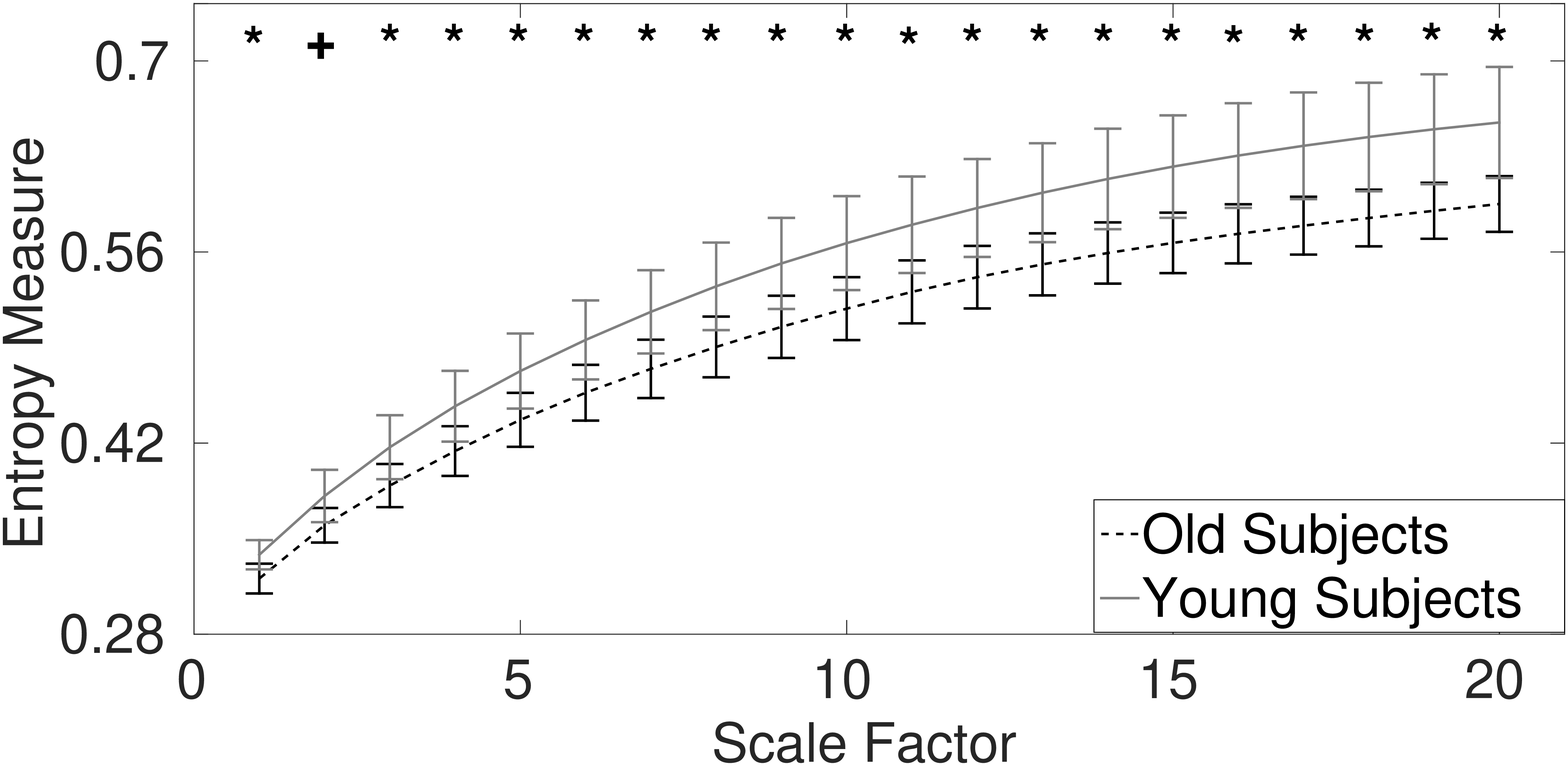}
\footnotesize{(d) MDE for Fantasia database}
\includegraphics[width=6.2cm,height=3.3cm]{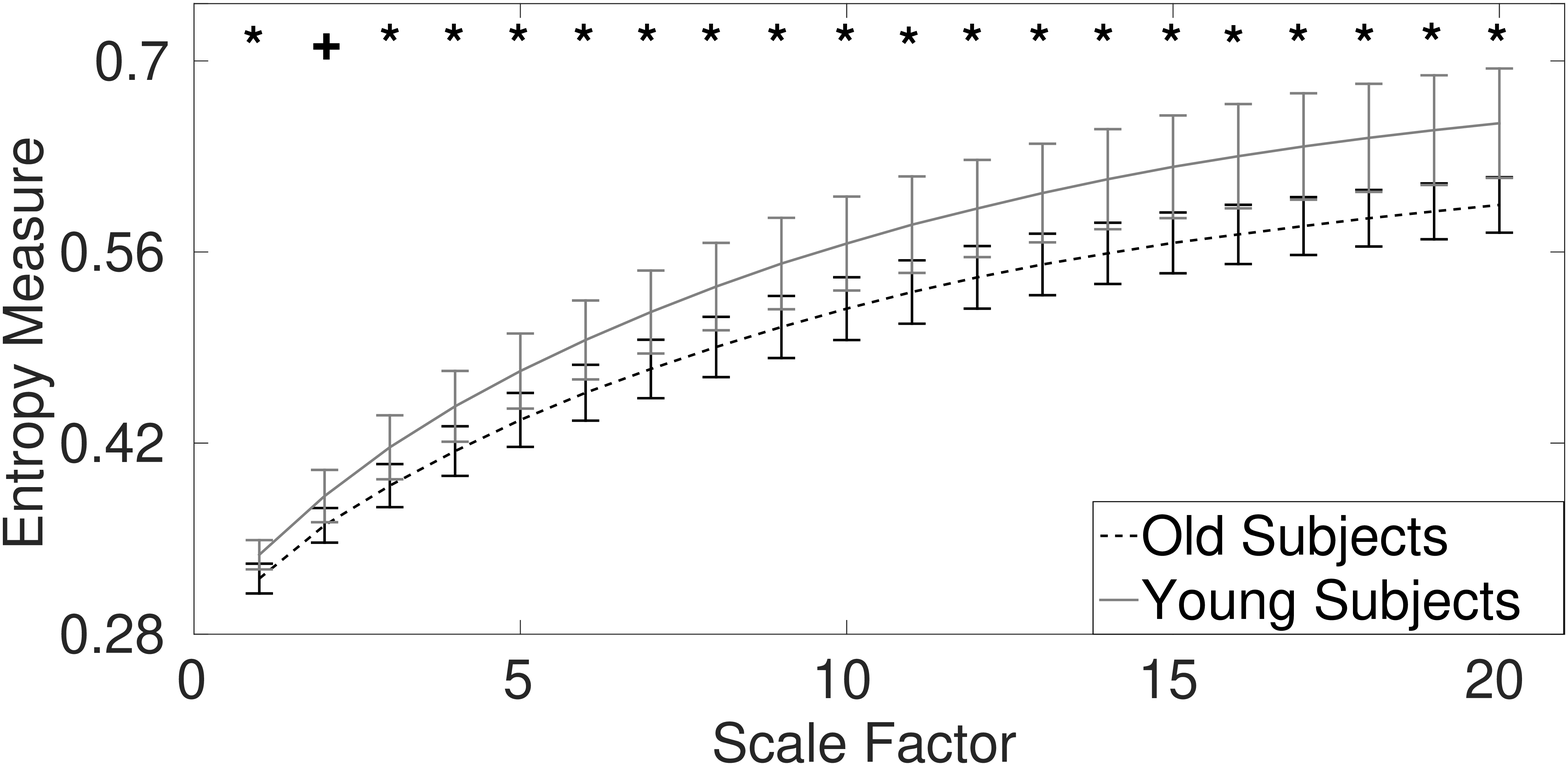}
\footnotesize{(e) RCMDE for Fantasia database}
\includegraphics[width=6.2cm,height=3.3cm]{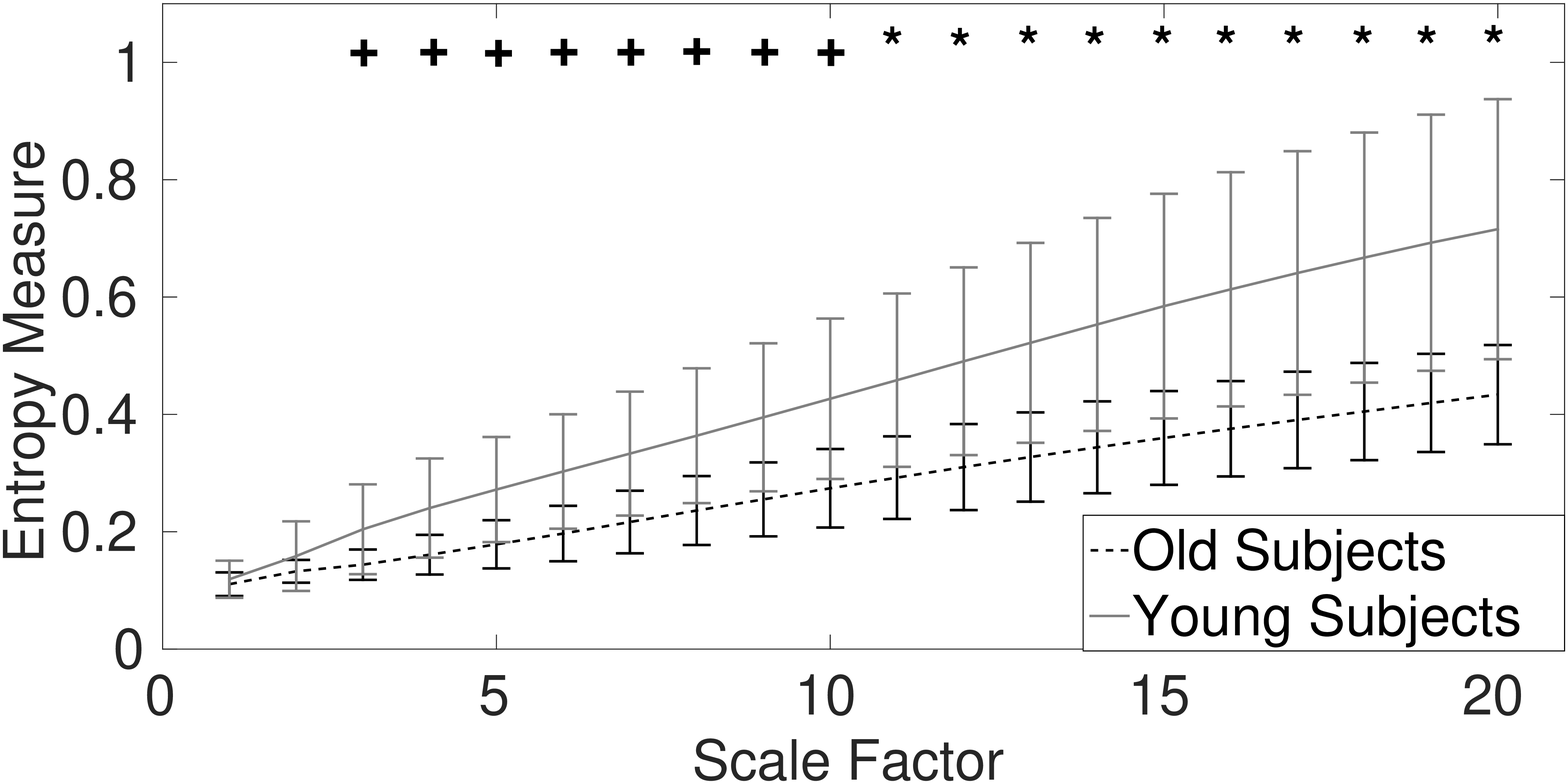}
\footnotesize{(f) RCMSE for Fantasia database}
\end{multicols}
			
\begin{multicols}{3}
\includegraphics[width=6.2cm,height=3.3cm]{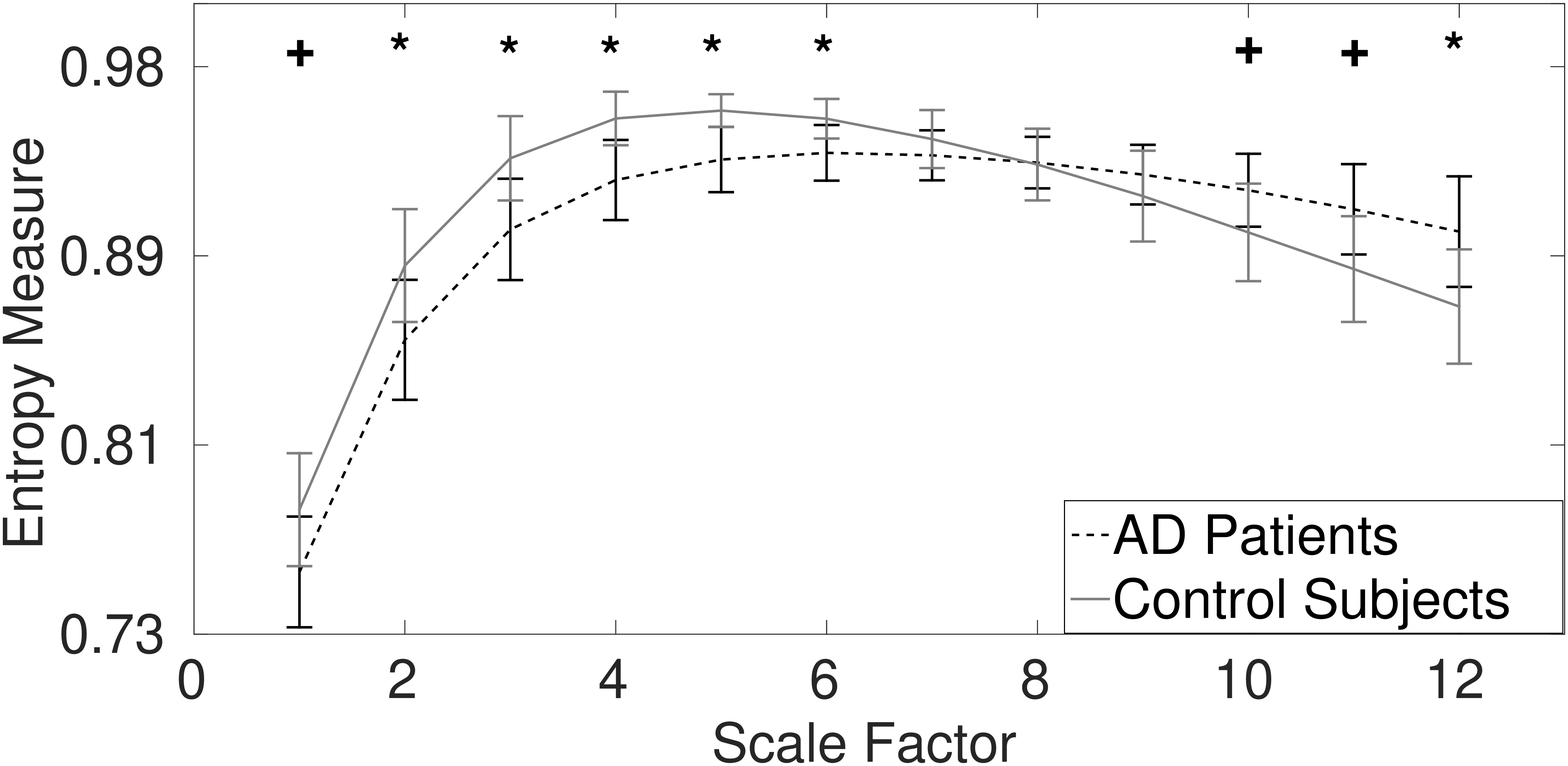}
\footnotesize{(g) MDE for AD subjects and controls}
\includegraphics[width=6.2cm,height=3.3cm]{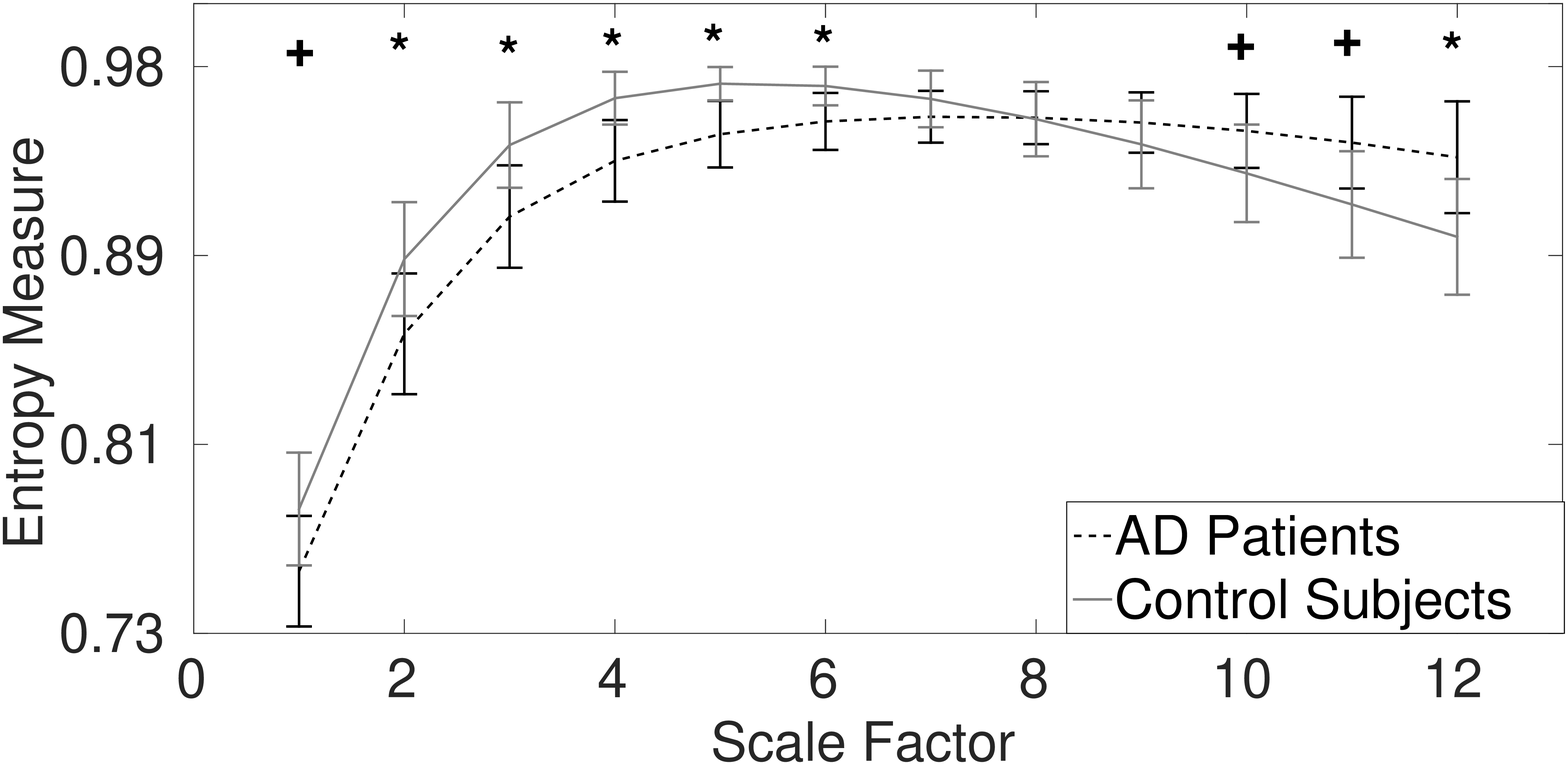}
\footnotesize{(h) RCMDE for AD subjects and controls}
\includegraphics[width=6.2cm,height=3.3cm]{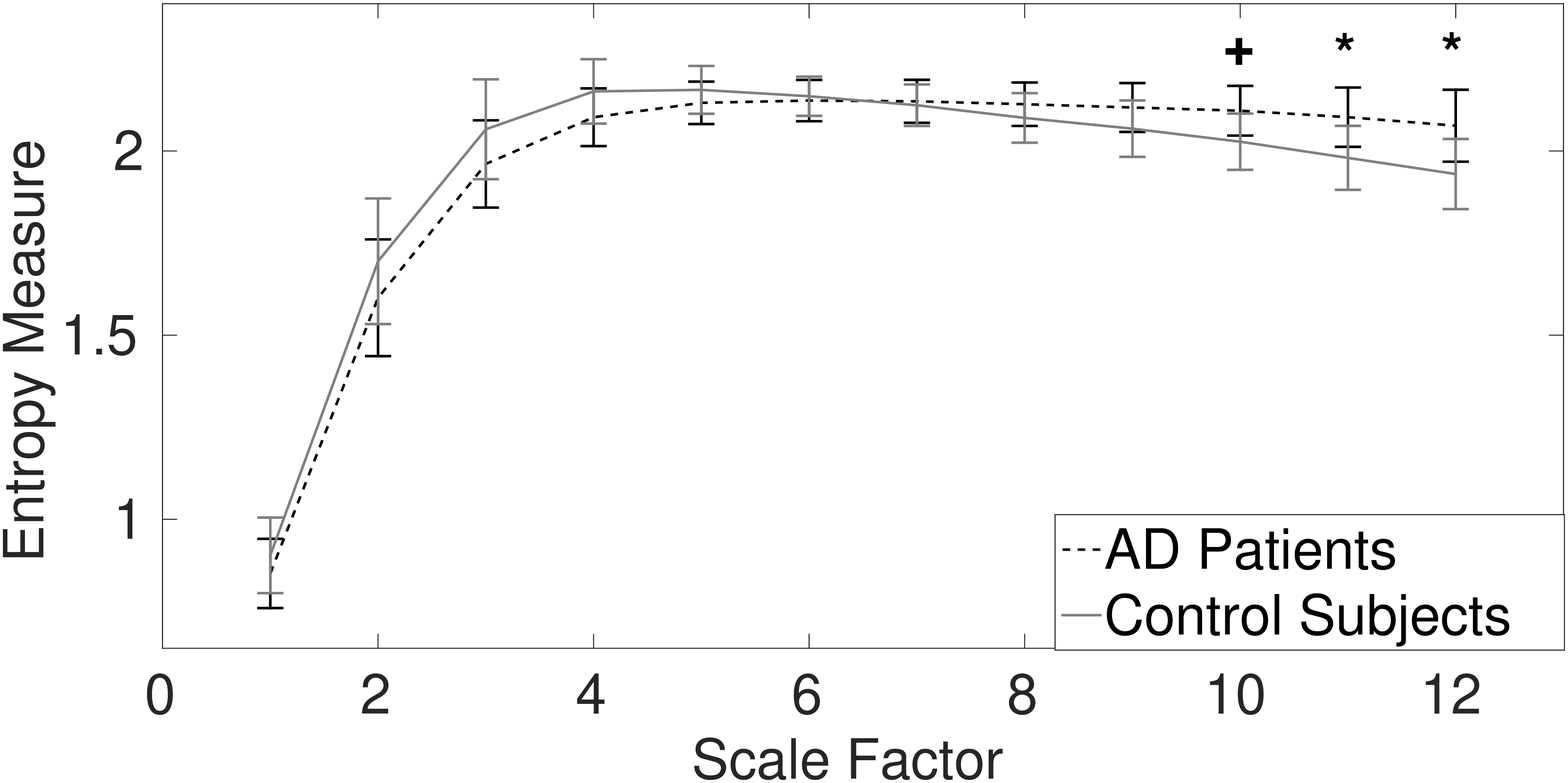}
\footnotesize{(i) RCMSE for AD subjects and controls}
\end{multicols}
\caption{ Mean value and SD of results of the MDE, RCMDE, and RCMSE computed from (a), (b), and (c) focal and non-focal EEGs, (d), (e), and (f) blood pressure recordings in Fantasia database, and (g), (h), and (i) resting-state EEGs activity in AD.}
\label{figurelabel}

\end{figure*}

\textit{2) Fantasia Dataset of Blood Pressure Recordings}: In Fig. 3(d)-(f), the error bars respectively show the mean and spread of the MDE, RCMDE, and RCMSE values computed from young and old subjects' blood pressure recordings in the Fantasia database. For each scale factor, the average of entropy values for elderly subjects are smaller than that for young ones using MDE, RCMDE, and RCMSE, in agreement with those obtained by the other entropy-based method \cite{nemati2013respiration}. For RCMDE, $\tau_{max}$ and $m$ respectively were 20 and 4. The computational time for the MDE, RCMDE and RCMSE results were about 1 hour, 5 hours, and 10 days, respectively. These considerable differences are due to the length of the signals (1,000,000 samples). For each scale factor and for each of MDE, RCMDE, and RCMSE, a Student's $ t $-test was also used to assess the statistical differences between the DisEn/SampEn values for young subjects versus elderly ones at all considered temporal scales. We adjusted the false discovery rate (FDR) independently for each entropy approach. The scales with the adjusted $ p $-values between 0.01 and 0.05 (significant) and less than 0.01 (very significant) are shown with + and *, respectively, in Fig. 3(d)-(f). The results show that the MDE and RCMDE lead to the very significant differences for elderly and young subjects at all scale factors, except the second scale, showing only significant difference. However, the RCMSE-based results do not show a significant difference at scales 1 and 2. The differences for scale factors 3-10 and 11-20 are significant and very significant, respectively. These facts show advantages of MDE and RCMDE over RCMSE.


\textit{3) Surface EEG Dataset of Brain Activity in AD}: We also analyze EEG signals from patients with AD and age-matched control subjects. The error bars showing the spread of the MDE, RCMDE, and RCMSE values are depicted in Fig. 3(g), (h), and (i), respectively. The average of MDE or RCMDE values for AD patients is smaller than that for controls at scale factors 1-7 (short-time scale factors), while after scale factor 8, the AD subjects' signals have larger entropy values (long-time scale factors). Similar results are obtained with RCMSE. All the results are in agreement with \cite{yang2013cognitive,labate2013entropic,escudero2015multiscale}. 
 
Moreover, a Student's $ t $-test was performed for AD patients' vs. controls' results. We adjusted the FDR independently for each of MDE, RCMDE, and RCMSE and each scale factor. As can be seen in Fig. 3(g)-(i), the results show that the MDE and RCMDE methods achieve very significant differences at scale factors 2, 3, 4, 5, 6, 11, and 12 and significant differences at scales 1 and 10, while the RCMSE leads to very significant differences at only scale factors 11 and 12 and a significant difference at scale factor 10. This suggests a better performance of the MDE and RCMDE approaches over the RCMSE algorithm for the distinction of EEG background activity related to AD.


The adjusted \textit{p}-values for EEGs in the AD dataset, unlike the previous two datasets, show that the AD patients' and controls' signals are not significantly different at several scale factors using MDE, RCMDE, and RCMSE, albeit MDE and RCMDE led to smaller adjusted \textit{p}-values. As an alternative approach, we integrate information from several temporal scales using slope values of complexity curves instead of considering entropy values separately at every time scale to distinguish different kinds of dynamics \cite{azami2017univariate}. The MSE and MDE profiles showing respectively the SampEn and DisEn values of each coarse-grained time series versus the scale factor were visually inspected to determine the range of scales over which the slope would need to be calculated. For example, the slopes of MDE and MSE profiles are used to discriminate controls and AD patients' surface EEGs. As can be seen in Fig. 3(g)-(i), for MDE profiles, the curves increase until scale factor 6. Then, the slope goes down and the DisEn values are nearly constant or decrease slightly. Therefore, we can divide each of the MDE curves into two segments: I) the first part corresponds to the steep increasing slope (small scale factors, i.e. $1\leq \tau \leq 6 $), and II) the second one contains the scale factors in which the slope of the DisEn values are smoother (large scale factors, i.e. $7\leq \tau \leq 12 $ ). Similarly for the MSE method, we divide the curves into two segments with scale factors $1\leq \tau \leq 4 $ and $5\leq \tau \leq 12 $. Note that the slope values of both parts were calculated based on the least-square approach.\\
Tabulation III shows the average $\pm$ SD of slope values of the MSE and MDE profiles for small and large time scales. The adjusted \textit{p}-values of the Student's \textit{t}-test were also calculated to investigate whether there is any significant difference between the AD and control groups. For small scale factors, no significant differences between both groups were found, whereas the differences between these groups were significant at large scale factors. More importantly, the adjusted \textit{p}-values obtained using MDE were noticeably smaller than those for MSE, showing that the MDE better discriminates the AD patients from controls than MSE. This again suggests that MDE is a better method than MSE to discriminate AD patients' from controls' signals.

\begin{table}
	\label{tab:table4}\caption{Average $\pm$ SD of slope values of the MDE and MSE profiles and adjusted \textit{p}-values for AD patients vs. controls over all channels and subjects.}
	\centering 
	\begin{tabular}{c*{5}{c}}
		Method &AD Patients& Controls& Adj. \textit{p}-values \\
		\hline
		MSE ($1\leq \tau \leq 4 $)     & 0.410$\pm$0.022& 0.418$\pm$0.024& 0.441\\
		MSE ($5\leq \tau \leq 12 $)    &0.002$\pm$0.019& -0.021$\pm$0.024&0.019\\
		MDE ($1\leq \tau \leq 6 $)    & 0.121$\pm$0.018& 0.111$\pm$0.025& 0.296\\
		MDE ($7\leq \tau \leq 12 $)   &-0.024$\pm$0.019&-0.053$\pm$0.015&0.001
	\end{tabular}
\end{table}

Overall, the results of the three real world datasets support the use of MDE and RCMDE over RCMSE due to its ability to produce similar complexity profiles, faster computational time and increased ability to find differences between physiological conditions. In future work, the ability of MDE and RCMDE to distinguish different kinds of dynamics of other physiological and non-physiological signals will be inspected. We also take into account the dynamics across the channels for multivariate time series \cite{ahmed2011multivariate} and will introduce multivariate MDE (mvMDE) and refined composite mvMDE.

\section{Conclusions}
To quantify the complexity of signals and to improve diagnostic or therapeutic interventions, we introduced and subsequently evaluated the MDE and RCMDE methods.
	
Our proposed methods MDE and RCMDE overcame the shortcomings of the existing multiscale entropy methods: 1) MSE and RCMSE values are undefined or unreliable for short signals, 2) MSE and RCMSE are not stable enough for short signals, and 3) the computation of MSE and RCMSE is not fast enough for some real-time applications. We evaluated the performance of MDE and RCMDE on several relevant synthetic signals and three datasets of real physiological signals. The results showed similar behavior in terms of complexity profiles of MSE or RCMSE and MDE or RCMDE although MDE and RCMDE are significantly faster, especially for long signals. Moreover, RCMDE was more stable than MDE for noisy signals, while for filtered biomedical signals, the performance of RCMDE and MDE was quite similar. For short signals, MDE and RCMDE, unlike MSE and RCMSE, did not lead to undefined values. In comparison with RCMSE, MDE and RCMDE discriminated better the elderly from young subjects and controls from AD patients, respectively, for Fantasia and AD datasets.

Overall, we expect MDE and RCMDE to play a prominent role in the evaluation of complexity in real signals.

\section*{Acknowledgment}

We would like to thank Dr Pedro Espino (Hospital Clinico San Carlos, Madrid, Spain) for his help in the recording of EEG signals and selection of epochs and to the Reviewers and Associate Editor for their feedback.

\ifCLASSOPTIONcaptionsoff
  \newpage
\fi
{
	\bibliographystyle{ieeetr}
	\bibliography{REFJMDE}
}

\end{document}